\documentclass[12pt,a4paper]{article}   
\usepackage[dvips]{graphicx}
\usepackage{amsmath,amsfonts,amssymb}  

\def\m@th{\mathsurround=0pt}

\def\fsquare(#1,#2){
\hbox{\vrule$\hskip-0.4pt\vcenter to #1{\normalbaselines\m@th
\hrule\vfil\hbox to #1{\hfill$\scriptstyle #2$\hfill}\vfil\hrule}$\hskip-0.4pt
\vrule}}
\def\Fsquare(#1,#2){
\hbox{\vrule$\hskip-0.4pt\vcenter to #1{\normalbaselines\m@th
\hrule\vfil\hbox to #1{\hfill$#2$\hfill}\vfil\hrule}$\hskip-0.4pt
\vrule}}

\def\addsquare(#1,#2){\hbox{$
	\dimen1=#1 \advance\dimen1 by -0.8pt
	\vcenter to #1{\hrule height0.4pt depth0.0pt%
	\hbox to #1{%
	\vbox to \dimen1{\vss%
	\hbox to \dimen1{\hss$\scriptstyle~#2~$\hss}%
	\vss}%
	\vrule width0.4pt}%
	\hrule height0.4pt depth0.0pt}$}}

\def\Addsquare(#1,#2){\hbox{$
	\dimen1=#1 \advance\dimen1 by -0.8pt
	\vcenter to #1{\hrule height0.4pt depth0.0pt%
	\hbox to #1{%
	\vbox to \dimen1{\vss%
	\hbox to \dimen1{\hss$~#2~$\hss}%
	\vss}%
	\vrule width0.4pt}%
	\hrule height0.4pt depth0.0pt}$}}
	
\def\Flect(#1,#2,#3){
\hbox{\vrule$\hskip-0.4pt\vcenter to #1{\normalbaselines\m@th
\hrule\vfil\hbox to #2{\hfill$#3$\hfill}\vfil\hrule}$\hskip-0.4pt
\vrule}}

\def\triple(#1){ \hbox{
   \normalbaselines\m@th\baselineskip0pt\offinterlineskip
   \vbox{ 
      \hbox{$\Flect(#1,#1,\hbox{ })$}\vskip-0.4pt
      \hbox{$\Flect(#1,#1,\hbox{ })$}\vskip-0.4pt
      \hbox{$\Flect(#1,#1,\hbox{ })$}\vskip-0.4pt
        }
      } }
	  
\newcommand{\np}{\nonumber\\}
\newcommand{\pn}{\par\noindent}

\newtheorem{thm}{Theorem}
\newtheorem{lem}{Lemma}
\newtheorem{prop}{Proposition}

\newtheorem{df}{Definition}
\newtheorem{cj}{Conjecture}
\def\dl{\frac{1}{6}}
\def\dle6{\frac{1}{4}}

\def\renx{\frac{x}{6}}
\def\ren6x{\frac{x}{4}}

\title{
Hidden E type structures in dilute A models
}
\author{ J. Suzuki\thanks{e-mail: sjsuzuk@ipc.shizuoka.ac.jp}\\
        \parbox{0.9\textwidth}{
        {\em
        \begin{center}
       Department of Physics, Faculty of Science\\
       Shizuoka University,\\
      Ohya 836, Shizuoka,\\
       Japan
        \end{center}
        }}
       }
\date{July 1999}
\begin{document}
\maketitle
\begin{abstract}
The hidden $E_{7} (E_{6})$ structure has been conjectured for the minimal
model ${\cal M}_{4,5} ( {\cal M}_{6,7}$) perturbed by
$\Phi_{1,2}$ in the context of conformal field theory(CFT).
Motivated by this, we examine the dilute $A_{4, 6}$ models, 
which are expected to be corresponding lattice models.
Thermodynamics of the equivalent one 
dimensional quantum systems is analyzed 
via the quantum transfer matrix approach.
Appropriate auxiliary functions, related to  kinks in the theory, 
play a role in
constructing functional relations among transfer matrices.
We successfully recover the universal $Y-$ systems 
and thereby Thermodynamic Bethe Ansatz equations
for $E_{6,7}$ from the dilute $A_{6,4}$ model,
respectively.
\end{abstract}
\clearpage

\section{Introduction}\label{intro}
The impact of perturbed conformal field theory (CFT) has many aspects
\cite{Zam891,Zam892}.
In this communication we explore one of its predictions; the 
"Trinity" among minimal unitary CFT theory  ${\cal M}_{p,p+1}, p=3,4, 6$ 
perturbed by $\Phi_{1,2}$,   lattice models off criticality,
(the Ising model in a field, the tricritical Ising model 
off the critical temperature and the tricritical 3 state Potts model) 
and the dilute $ A_{3,4,6}$ models.
Many results have already been accumulated on the equivalence in 
universality
\cite{HenkSal}-\cite{MO97}. 
The scaling exponents of the dilute $ A_{3}$ model in
periodic or open boundary conditions have been evaluated analytically
\cite{WNS1}-\cite{BS2}.
They agree with  numerical results for the Ising model in a 
magnetic field.
Masses for eight elementary excitations of the dilute $ A_{3}$ model 
are found to be proportional to components of 
the (largest) eigenvector of Cartan matrix
for $E_8$\cite{BS1, MO97}.
Vertex operators of $A^{(2)}_2$, which is the symmetry of
the dilute $ A_{3}$ model at criticality,
satisfy a set of relations indicating the hidden
$E_8$ structure \cite{Hara}.

Especially, we like to call attention to 
 thermodynamics of a 1D system related to the dilute $ A_{3}$ model
in \cite{BWN}.
A set of solutions to the eigenvalue problem of the 1D Hamiltonian has been
identified in exquisite "string" forms \cite{BWN, GN1, GN2}.
Nine of them are expected to
contribute nontrivially in the thermodynamic limit.
This observation leads to a set of integral equations (Thermodynamic Bethe ansatz
, TBA ) which determines the free energy.
Remarkably, TBA exhibits the underlying $E_8$ structure 
\cite{BazResProg}.

In \cite{SuzE8}, we have attacked the same problem in a different setting.
By following general frameworks, one represents the free energy
by the largest eigenvalue of the "quantum transfer matrix"
(QTM) acting on a virtual space \cite{Suz85} -\cite{DeVe92}. 
We have managed to solve the (single) eigenvalue problem of commuting QTM
by introducing 
auxiliary functions related to fusion of QTM
\cite{Klu92,JKSfusion, KSS98}.
We will simply call them fusion QTMs.
Eight fusion QTMs  are found to 
 satisfy a closed set of functional relations related to $E_8$.
A quantum analogue of the Jacobi-Trudi formula \cite{BR, KOS}, as well as 
 combinatorial aspects in terms of 
"Yangian analogue" of Young tableaux \cite{BR,SuzG2, KS, KOS, Tsuboi98}
play a fundamental part in the proof of the relations.
Nice analytic properties of fusion QTMs
allow for the transformation of functional relations into 
coupled integral equations.
The resultant thermodynamic Bethe ansatz (TBA) equation 
yields a direct  evaluation of free energy. 
Again it coincides with a hypothetical TBA for
the $E_8$ theory.

As promised in \cite{SuzE8},
 we carry out this program for the dilute $A_4$ and $A_6$ models.
A novel feature lies in the fact that "a box" of the "Young tableaux"
is no longer the fundamental constituent.
This may be natural in view of the representation theory;
the vector representation is no longer minimal.
In the language of the S-matrix theory,  boxes present
breathers, rather than kinks.
We conjecture  explicit forms of QTMs related to these kinks.
The investigation on these kink QTMs reveals  connections
between $U_q(A^{(2)}_2)$ modules of symmetric tensors of the vector
representation and $U_q(\widehat{E_6}), U_q(\widehat{E_7})$ modules
 when $q$ equals to proper root of unity.
With help of this observation,  closed functional relations
among fusion QTMs are also found for the dilute $A_{4, 6}$ models.
Quite parallel to the dilute $A_{3}$ model, one recovers TBAs
expected for $E_{6, 7}$ \cite{BazResProg}.

The paper is organized as follows.
As the subject may not be so familiar to readers, 
we present a brief survey on the QTM approach, together with
the sketch of the idea of analytic Bethe ansatz in section \ref{survey}.
The dilute $A_L$ model is briefly described in section \ref{review-model}.
The $sl_3$ fusion structure of the model is 
presented in view of  analytic Bethe ansatz, 
the "Yangian analogue" of Young tableaux and the quantum Jacobi-Trudi formula
in section \ref{review-qJT}.
We concentrate on the dilute $A_4$ model in sections \ref{kinkQTM} and 
\ref{Tsys-A4}.
The explicit form of QTM related to the "kink" is proposed in section \ref{kinkQTM}.
Yangian homomorphisms among $Y(E_7)$ modules serve as 
a useful guide in search of the form.
The rest of QTMs are defined and their functional relations are examined in
section \ref{Tsys-A4}.
The result coincides with prediction in \cite{KNS1}.
Similar results for the dilute $A_6$ model are given in section \ref{summary-A6}.
With piece of information on analyticity of these QTMs, supported by numerics,
the desired  TBAs are recovered in section \ref{Ysys}.
We conclude the paper with a short summary in section \ref{conclusion}.


\section{Survey on the QTM approach}\label{survey}

%
Exact evaluation of physical quantities at finite temperatures poses
serious difficulties even for integrable models.
One has to go much beyond mere diagonalization of a Hamiltonian;
summation over eigenspectra must be performed.

The string hypothesis brought the first breakthrough and success.
It postulates dominant solutions to the Bethe ansatz equation (BAE) in the
thermodynamic limit.
In a sense, the method tackles the combinatorial aspect of the problem  
directly.
%
%
\subsection{QTM }

The quantum transfer matrix (QTM) method takes a different route.
It utilizes the famous mapping between the Hamiltonian ${\cal H}_M$ of 
a 1D quantum system and the row to row transfer matrix  
$T_{\footnotesize{RTR}}(u)$ of the corresponding 2D classical model
\cite{Suz85} -\cite{DeVe92}.
In the present context, the latter is given by
$$
(T_{\hbox{\footnotesize{RTR}}}(u))^{\{b\}}_{\{a\}} =\prod_{j=1}^M 
\raise 2mm \vtop{\hbox{$b_j$}\hbox{$a_j$}}
\> \framebox[0.4cm][c]{$u$} \>
\raise 2mm \vtop{\hbox{$b_{j+1}$}\hbox{$a_{j+1}$}}.
$$
Here the box represents the RSOS weight and $M$ is
the number of sites.
See appendix for explicit weights for the dilute $A_L$ models.
The parameter $u$ represents the anisotropy of interactions 
between horizontal and vertical directions, and is called the 
spectral parameter.
The explicit relation between  ${\cal H}_M$ and  
$T_{\footnotesize{RTR}}(u)$ reads,
$$
T_{\footnotesize{RTR}}(u) 
\sim T_{\footnotesize{RTR}}(0) (1+ \frac{u}{\epsilon} {\cal H}_M +O(u^2)),
$$
where $\epsilon$ is the normalization parameter of the Hamiltonian.
The essential idea in the QTM approach is encoded in the following identity,
\begin{eqnarray*}
\exp (-\beta {\cal H}_M) &= &
\lim_{N \rightarrow \infty} 
(1+ \frac{u}{\epsilon}  {\cal H}_M )^N |_{u \rightarrow -\beta \epsilon/N} \\
&=&\lim_{N \rightarrow \infty}  
\bigr ( T'_{\footnotesize{RTR}}(u=-\beta \epsilon/N)\bigl)^N  \\
T'_{\footnotesize{RTR}}(u) &:=&
 \Bigl( T_{\footnotesize{RTR}}(u) ( T_{\footnotesize{RTR}}(0))^{-1}
   \Bigr )^N.
\end{eqnarray*}
Namely, the partition function of the original problem is transformed into
that of the 2D classical models on $M \times N$ sites.
The fictitious dimension $N$ is sometimes referred to as the
Trotter number.
We can interpret  $T'_{\footnotesize{RTR}}(u)$ as
the row to row transfer matrix in the "vertical" direction.
Similarly, one can construct a transfer matrix propagating in the 
"horizontal" direction $T'_{\footnotesize{QTM}}(u)$, which acts on
$N$ sites.
Thereby we have 
\begin{eqnarray*}
 {\hbox{Tr }} \exp (-\beta {\cal H}_M) =
\lim_{N \rightarrow \infty} {\hbox{Tr }}
 T'_{\footnotesize{QTM}}(u=-\beta \epsilon/N) ^M .
\end{eqnarray*}

It may be better to rewrite this in the form,

\begin{eqnarray*}
\lim_{M \rightarrow \infty} \frac{1}{M} \log 
   \Bigl (  {\hbox{Tr }} \exp (-\beta {\cal H}_M)  \Bigr )  =
\lim_{N \rightarrow \infty} \lim_{M \rightarrow \infty} 
\frac{1}{M}  \Bigl (   {\hbox{Tr }}
          T'_{\footnotesize{QTM}}(u=-\beta \epsilon/N) ^M  \Bigr )  .
\end{eqnarray*}
The exchangeability of two limits are proven in \cite{Suz85}. \pn
The gap opens up 
between the largest and the second largest eigenvalues of
$T'_{\footnotesize{QTM}}(u)$.
In the thermodynamic limit $M \rightarrow \infty$, 
we only have to deal with the largest eigenvalue of QTM.
This strongly contrasts to the spectra of $T'_{\footnotesize{RTR}}(u)$.
One observes almost degenerate low lying excitations in the latter case
as $M \rightarrow \infty$.
The evaluation of free energy per site of the 1D quantum system 
is thus reduced to the largest eigenvalue problem of $T'_{\footnotesize{QTM}}(u)$.
We are free from the summation problem.

This is, unfortunately, not the happy end of the story.
The Trotter number should be sent infinity
at the end.
The diagonalization of QTM is accomplished by the application
of the Bethe ansatz method.  
The BAE depends nontrivially on $N$, which originates from
the local interaction parameter $u$.
Thus we can not retort to the simple-mind
application of the 
usual scheme of converting the transcendental equation
into the integral equation.
This makes the extrapolation  $N \rightarrow \infty$ quite
nontrivial.
%
%
\subsection{commuting QTM}
Instead of dealing with  BAE roots directly, 
we employ a different idea.
The integrable structure of  the underlying model 
allows for the introduction of 
 one parameter family of commuting QTMs which is labeled by a novel complex
 parameter $x$.
For the explicit demonstration of this,
 we adopt a more sophisticate approach \cite{Klu92, Klu93} 
 than the one presented above.
One introduces  a "staggered manner"  QTM 
to avoid $( T_{\footnotesize{RTR}}(0))^{-1}$ factor in the definition of 
$T'_{\footnotesize{RTR}}(u)$.

$$
(T_{\hbox{\footnotesize{QTM}}}(u,x))^{\{b\}}_{\{a\}} =\prod_{j=1}^{N/2}
\mbox{\parbox[c][0.9cm]{0cm}{}}^{b_{2j-1}}_{a_{2j-1}}
\fbox{\parbox[c][0.7cm]{0.7cm}{ {\scriptsize $u\!+\! ix$}  }} \>
\mbox{\parbox[c][0.9cm]{0cm}{}}^{b_{2j}}_{a_{2j}}
\> \>
\mbox{\parbox[c][0.9cm]{0cm}{}}^{\,\,a_{2j}}_{a_{2j+1}}
\fbox{\parbox[c][0.7cm]{0.7cm}{ {\scriptsize $u\!-\! ix$}  }} \>
\mbox{\parbox[c][0.9cm]{0cm}{}}^{b_{2j}}_{b_{2j+1}}.
$$
The following relation is still valid,
\begin{eqnarray*}
\beta f &=&
-\lim_{M\rightarrow \infty} \frac{1}{M} 
   \ln {\hbox{Tr }} \exp(-\beta {\cal H}_{\epsilon})  \\
 &=& -\lim_{N\rightarrow \infty}\ln  \Bigl( \hbox{the largest eigenvalue of }
 T_{\hbox{\footnotesize{QTM}}}(u= -\epsilon \frac{\beta}{N}, x=0) \Bigr ).
\end{eqnarray*}

As is emphasized above, we find the
intriguing fact, commutativity of QTMs,
$$
[T_{\hbox{\footnotesize{QTM}}}(u,x), T_{\hbox{\footnotesize{QTM}}}(u,x') ]=0.
$$

Generally, one can construct a "higher spin" QTM 
$T_{\hbox{\footnotesize{fusion}}}(u,x)$ by the fusion procedure.
The Yang Baxter integrability also assures 
commutativity among these generalized 
QTMs,
$$
[T_{\hbox{\footnotesize{fusion}}}(u,x), 
T_{\hbox{\footnotesize{fusion'}}}(u,x') ]=0.
$$
Note that the factor $u$ is in common.
Hereafter we will sometimes drop this common factor in commuting QTMs.
We also sometimes use same notations  for transfer matrices and
their eigenvalues
as we are considering them on the identical eigenspace.

We utilize the existence of  the complex $x$ plane in
which QTMs are simultaneously diagonalizable.
There exist functional relations among these fusion QTMs
in the $x$ plane.
Our idea is to utilize these functional relations in place of BAE.
See \cite{BP, KP} for the discussion in the case of
usual row to row transfer matrices.
Our motivation is simple. 
The number of roots is of order  $N$, the Trotter number.
All these locations  are changing
with  $N$, while functional relations depend on $N$ weakly.
The dependence can 
be summarized in the known scalar factors in  functional relations.
One may expect  the tractable limit  $N \rightarrow \infty$ 
for functional relations. 
The problem of combinatorics (summation over eigenspectra ) is then 
reduced to the study of functional relations 
\footnote{
One can adopt different auxiliary functions
from the fusion hierarchy.
Results on other choices of auxiliary functions
 for several models, see \cite{Klu93},\cite{KluBa95}-\cite{SSSU}}
and analytic structures of fusion QTMs, as  will be discussed soon below.


\subsection{Analytic Bethe ansatz and functional relations}\label{ABA}

The QTMs should not possess singularities in the $x$ plane
as Boltzmann weights are regular functions of $x$.
BAE can be interpreted as the pole-free condition
of QTMs in the complex $x$ plane.
Conversely, the analyticity requirement imposes restrictions on
 explicit eigenvalues of QTMs.

The analytic Bethe ansatz was proposed in \cite{Reshet1}, 
as a tool in deriving  expressions of 
eigenvalues of transfer matrices.
It starts from a simple observation; the eigenvalue at
the "vacuum sector" is determined by diagonal elements of 
$R$ matrix, which are referred to as vacuum expectation values.
In general sectors, eigenvalues should be modified such that each 
vacuum expectation value is "dressed" by appropriate
combination of Baxter's $Q$ operators.
See eq.(\ref{t1expr}) for a typical example.
The combination is determined by requiring analyticity of
the transfer matrix.
We call the resultant expression, Dressed Vacuum Form (DVF).

A "universal" BAE has been proposed in \cite{Reshet1, Reshet2};
one can write down BAE for the model based on $U_q(\hat{\mathfrak{g}})$ 
using only algebraic data of $U_q(\mathfrak{g})$.
Starting from a properly chosen "highest weight term", 
we can construct a pole-free set of functions under BAE.
In \cite{KS} a similarity  has been pointed out
between the above procedure 
and the construction of the highest weight module of Lie algebra.
It leads to an assumption that there exists a set of 
functions, pole-free under BAE, 
corresponding to an irreducible module
of quantum affine Lie algebra.
The set is naturally identified with the eigenvalue of the transfer matrix
of which trace is taken over the irreducible module.
This has been promoted as an axiom in \cite{KS} and subsequent papers 
\cite{ KOS, Tsuboi98} producing fruitful results.
We take  $sl_2$ as the simplest example. 
By $V_m(x)$ ($T_m(x)$), we mean the m+1-dimensional $sl_2$ module, and
the associated transfer matrix.
The DVF of $T_1(x)$ consists of two terms. 
We do not specify their forms and represent them by 
boxes with letters 1 and 2,
$$
T_1(x)= \Fsquare(0.5cm,1 )_{x} + \Fsquare(0.5cm,2 )_{x}.
$$
Each box carries spurious poles, which are actually canceled due to
BAE.
We represent this situation graphically as,
$$
\Fsquare(0.5cm,1 )_{x} \rightarrow \Fsquare(0.5cm,2 )_{x}.
$$
The eigenvalue of a fusion QTM can be apparently represented by
sum over products of "boxes" with various letters and spectral parameters.
For example, one can construct a transfer matrix 
of which auxiliary space acts on the symmetric subspace of 
$V_1 \times V_1$. 
We associate to this the set of glued boxes,
$
\Fsquare(0.5cm,i_1 )_{\,x-i} \,
\Fsquare(0.5cm,i_2 )_{\,x+i}, (i_1\le i_2)$.
The difference in spectral parameters is fixed so as 
to match the singularity
of $R$ matrix.
The cancellation of spurious singularities is again depicted as
$$
\Fsquare(0.5cm, 1) \Fsquare(0.5cm, 1) \rightarrow
\Fsquare(0.5cm, 1) \Fsquare(0.5cm, 2) \rightarrow
\Fsquare(0.5cm, 2) \Fsquare(0.5cm, 2),
$$
where we omit spectral parameters.
The eigenvalue of $T_2(x)$ is given by the sum of three diagrams in
the above.
Extension to general $T_m$ is now obvious.
Starting from the "highest weight" term, 
$\begin{tabular}{|l|l|l|l|}
\hline
1& 1& $\cdots$& 1 \\
\hline
\end{tabular}, 
$ 
we  have a  cancellation diagram,

$$
\begin{tabular}{|l|l|l|l|}
\hline
1& 1& $\cdots$& 1 \\
\hline
\end{tabular} 
\rightarrow 
\begin{tabular}{|l|l|l|l|}
\hline
1& 1& $\cdots$& 2 \\
\hline
\end{tabular}
\rightarrow  \cdots \rightarrow
\begin{tabular}{|l|l|l|l|}
\hline
2& 2& $\cdots$& 2 \\
\hline 
\end{tabular}.
$$
The sum over them, regarded as expressions,  yields the eigenvalue of $T_m(x)$.
One can easily identified the diagram with the 
crystal graph of the m-fold tensor of $U_q(sl_2)$ representing
the irreducible module.
%

Suppose that we have a short exact sequence 
among tensor products of irreducible modules
of quantum affine Lie algebra,
$$
0 \rightarrow W_0 \otimes W_1 \rightarrow W_2 \otimes W_3
 \rightarrow W_4 \otimes W_5 \rightarrow 0,
 $$
then a functional relation follows,
 $$0 = T_{W_0} T_{W_1} - T_{W_2} T_{W_3} + T_{W_4} T_{W_5}. $$
Remark that spectral parameter dependencies are implicit in  $W_i$s
\footnote{To be precise, we first consider a vertex model of 
which quantum space (auxiliary space)
is given by  $W_i$  and denote the transfer matrix  by
 $T_{W_i}$. 
 Later section we use the same notation for the transfer matrix of the
 corresponding RSOS model  }.
The desired functional relations are derived as a consequence of
 relations among affine modules.
Even if exact sequences are not available,
one can still check the validity of hypothetical
functional relations using explicit forms of transfer matrices,
which can be derived by applying the analytic Bethe ansatz.
Indeed,  functional relations for $sl_2$ are easily 
derived without knowledge on exact sequences.
By using the above box-representation, one can derive 
 graphically, 
\begin{equation}
T_m(x-i) T_m(x+i) = g_m(x)+ T_{m-1}(x) T_{m+1}(x)  \quad m\ge 1, 
\label{tfor2}
\end{equation}
where $g_m(x)$ is a known scalar function which depends on $N$.

Such functional relations are sometimes referred to as
the $T-$ system.
%
%
\subsection{Functional relations and Thermodynamic Bethe Ansatz}

Unfortunately, functional relations alone do not provide enough
information on the explicit eigenvalues.
This can be easily seen from the fact that  excited states' eigenvalues 
satisfy same algebraic relations.
One needs additional information on analyticities of fusion
QTMs.
Let us  again demonstrate this for the ${sl}_2$ case.
One conveniently rewrites the $T-$ system (\ref{tfor2}) in terms of
$Y_m(x)=T_{m-1}(x) T_{m+1}(x)/ g_m(x)$,
\begin{equation}
 Y_m(x-i) Y_m(x+i) = (1+Y_{m-1}(x))(1+Y_{m+1}(x)),
\label{sl2Y}
\end{equation}
where a known property of the $g-$ function, 
$g_m(x+i)g_m(x-i)=g_{m-1}(x)g_{m+1}(x)$ is used.

Consider functions in the largest eigenvalue sector of $T_1(x)$.
We have convincing numerical evidences for the conjecture 
that zeros of  $T_m(x)$ approximately lie on the curve 
$\Im x \sim \pm(m+1) $. 
Then both sides of (\ref{sl2Y}) are analytic, nonzero
and asymptotically constant (ANZC) within the 
strip $\Im x \in [-1,1]$.
Strictly speaking, we must modify the lhs for $m=1$. 
We will not go into such detail in this introductory part. 

This piece of information is now sufficient to transform
the algebraic relations to integral equations which enable
the explicit evaluation of $Y_m$ and then $T_m$.

Take the logarithmic derivatives of both sides of (\ref{sl2Y})
and perform Fourier transformations.
Let  $\tilde{dl}Y_m[k]$ be the  Fourier transformation
of the logarithmic derivative of $Y_m(x)$.
Thanks to the ANZC property, the Cauchy theorem applies.
The resultant equation is simply given by 
$$
2 \cosh k \tilde{dl}Y_m[k] = 
  \tilde{dl}(1+Y_{m-1})[k]+ \tilde{dl}(1+Y_{m+1})[k]. (m \ge 2)
$$
Remarkably, both sides 
only contain functions with the same Fourier mode.
Dividing both sides by $ 2 \cosh k $, performing 
inverse Fourier transformation and integrating once over $x$,
we reach the integral equation, which is identical to
TBA,
\begin{equation}
\log Y_m (x) = \int_{-\infty}^{\infty} K(x-y) 
\log(1+Y_{m-1})(1+Y_{m+1})(y) dy. \qquad (m \ge 2)
\label{sl2TBA}
\end{equation}
$K(x)$ denotes the Fourier transformation of $1/2 \cosh k$.
Though we have omitted above, the rhs of the equation  
(\ref{sl2TBA}) for $m=1$ has a nontrivial scalar factor
originated from $g$ function.
It thus brings a $N-$ dependency, however, by the 
combination $uN$ in the $N \rightarrow \infty$ limit.
Remembering that u is inversely proportional to
$N$, we can send $N \rightarrow \infty$ analytically!
The resultant drive term depends only on $\beta$.

In this way, we take a completely different route from the string hypothesis
but reach the same conclusion.
In the absence of appropriate  conjectures on dominant patterns
of roots, our method has an explicit advantage in attacking the problem.  
This is the case with the dilute $A_{4,6}$ models. 
\footnote{Although yet unpublished,
 there is also progress in view of string hypothesis for these cases.
The author thanks V.V. Bazhanov and O. Warnaar for information.}

In the rest of this paper, we shall extensively apply the above ideas
to these cases.
Before that, we repeat lessons from the above.
To find functional relations is not enough.
One must find them having ANZC property in appropriate domains
of the complex $x$ plane.
This is the most crucial step in the present approach.
%
%
\section{The dilute $A_L $ model}\label{review-model}
The dilute $A_L$ model is proposed in \cite{WNS1, WNS2} as
an elliptic extension of the Izergin-Korepin model \cite{IK}.
(See  \cite{Kuniba} for an elliptic extension of the different type.)
The model is of the restricted SOS type with local 
variables $\in \{1,2,\cdots,L \}$.
The variables $\{a, b\} $ on neighboring sites
should satisfy adjacency condition, $|a-b|\le 1$.
The solvable weights contain parameters $u, q$ and $\lambda$.
We supplement their explicit forms in appendix.
The model exhibits four different physical regimes depending on parameters,
\begin{itemize}
 \item regime 1. $0<u<3 , \,\lambda=\frac{\pi L}{4(L+1)},\, L\ge 2 $
 \item regime 2. $0<u<3, \, \lambda=\frac{\pi (L+2)}{4(L+1)},\,  L \ge 3$
 \item regime 3. $3-\frac{\pi}{\lambda}<u<0, \,
  \lambda=\frac{\pi (L+2)}{4(L+1)}, \, L \ge 3 $
 \item regime 4.  $3-\frac{\pi}{\lambda}<u<0,\, 
  \lambda=\frac{\pi L}{4(L+1)}, \, L \ge 2$
\end{itemize}
We are interested in  regimes 2 and 3.
As in section\ref{survey}, one defines the Hamiltonian of 
the associated 1D quantum chain by
$$
{\cal H}_{\epsilon} = \epsilon \frac{\partial}{\partial u} 
\ln T_{\hbox{\footnotesize{RTR}}}(u) |_{u=0}
$$
as in \cite{BWN}. 
$\epsilon=-1,(1)$ corresponds to regimes 2 (3), respectively.

The one particle excitations for the dilute $A_L$ case have  been
examined in \cite{SB3, SB4} for $L=$3,4 and 6. (See also  \cite{MO97}
for another derivation for $L=3$) 
Eight, seven and six particles are identified respectively, and their masses 
 are summarized by a single formula in the trigonometric limit,
$$
m_j \sim  \sum_a \sin(\frac{a \pi}{g^{\vee}}),
$$
where $g^{\vee}$ is 30, 18,  12 for $L=$3,4, 6  and 
is  nothing  but 
the dual Coxeter number for $E_8, E_7 $
and $E_6$, respectively.
We present sets of allowed $a$'s 
in table 1,2 for
$L=$4, 6, which are of our current interest.

\begin{center}
\begin{tabular}{|l|l|}
\hline
$j$ &    set of allowed $a$'s for $A_4$\\
\hline 
1 (2) &    $\{1,7 \}$ \\
2 (5) &    $\{2,6,8 \}$ \\
3 (7)  &    $\{3,5,7,9\}$ \\
4 (6)&    $\{4,6,8\}$ \\
5 (4)  &    $\{5,7\}$ \\
6  (1)&    $\{6\}$  \\
7  (3) &    $\{4,8 \} $  \\
\hline
\end{tabular}  
\end{center}         
\begin{center}
Table1
\end {center}

\phantom{ }

\begin{center}
\begin{tabular}{|l|l|}
\hline
$j$ &    set of allowed $a$'s for $A_6$\\
\hline 
1, 5 (1) &    $\{4 \}$ \\
2 ,4 (4) &    $\{3,5 \}$ \\
6 (3)  &    $\{1,5\}$ \\
3  (6)&    $\{2,4,6\}$  \\
\hline
\end{tabular}           
\end{center}         
\begin{center}
Table2
\end {center}

A number $k$ in the bracket means that it corresponds to the $k-$ th
light particle.
Note that  vectors of
the form $(m_1, m_2,\cdots)$, coincide with the eigenvectors
of Cartan matrices for $E_7$ and $E_6$, respectively.
The exponents $\{a\}$ will re-appear in a novel context later.
The leftmost numbers, which are just indices up to the present, will be 
 connected to indices for nodes in the Dynkin diagrams of  $E_7$ and $E_6$.
 \begin{figure}[hbtp]
\centering
  \includegraphics[width=8cm]{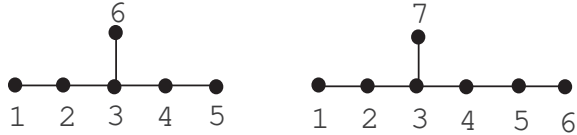}
\caption{Dynkin diagrams for $E_7$ and $E_6$.}
\label{dynkin}
\end{figure}
%
\section{$sl_3$ fusion structure and
 Quantum Jacobi Trudi formula}\label{review-qJT}

The $sl_3$ type fusion structure in the dilute $A_L$ model
has been discussed in \cite{GPZ}.
This comes from the singularity of RSOS weights at
$u=\pm 3$; the face operator becomes a projector related to
 $sl_3$ at these points.
One picks up a desired subspace from tensor products
of spaces using these projectors.
The adjacency conditions of local states are described  by
combinatorics of tableaux.

We are interested in eigenvalues of fusion QTMs.
Then the most relevant is the fact that 
these eigenvalues are again expressible in terms of 
"Young tableaux" depending on spectral parameters, as exemplified 
in section \ref{ABA}
for the $sl_2$ case.

Explicitly, the eigenvalue $T_1(u,x)$ of $T_{\hbox{\footnotesize{QTM}}}(u,x)$
is given by
\begin{eqnarray}
T_1(u,x)&=& 
w \phi(x+\frac{3}{2}i) \phi(x+\frac{1}{2}i)\frac{Q(x-5/2 i)}{Q(x-1/2 i)} \np
&+&\phi(x+\frac{3}{2}i) \phi(x-\frac{3}{2}i) 
\frac{Q(x-3/2i)\, Q(x+3/2 i)}{Q(x-1/2 i)\, Q(x+1/2 i)} \np
&+ &w^{-1} \phi(x-\frac{3}{2}i) \phi(x-\frac{1}{2}i)  
 \frac{Q(x+5/2 i)}{Q(x+1/2 i)},  \label{t1expr} \np
Q(x)&:=&\prod_{j=1}^{N} h[x-x_j] \np
\phi(x) &:=& \Bigl(\frac{h[x+(3/2-u)i] h[x-(3/2-u)i]}{h[2i] h[3i]}\Bigr)^{N/2},
\qquad h[x] := \theta_1(ix),  \np
\end{eqnarray}
and $w=\exp(i \pi \frac{\ell}{L+1})$ where  
$\ell=1$ for the largest eigenvalue sector.
$\theta_1(ix)$ is defined in  appendix.
The parameters, $\{ x_j \}$ are solutions to BAE,
\begin{equation}
w \frac{\phi(x_j+i)}{ \phi(x_j-i)}=\frac{Q(x_j-i) Q(x_j+2 i)}{Q(x_j+i)Q(x_j-2 i)},
\quad j=1,\cdots, N.
\label{bae}
\end{equation}

\begin{figure}[hbtp]
\centering
  \includegraphics[width=4cm]{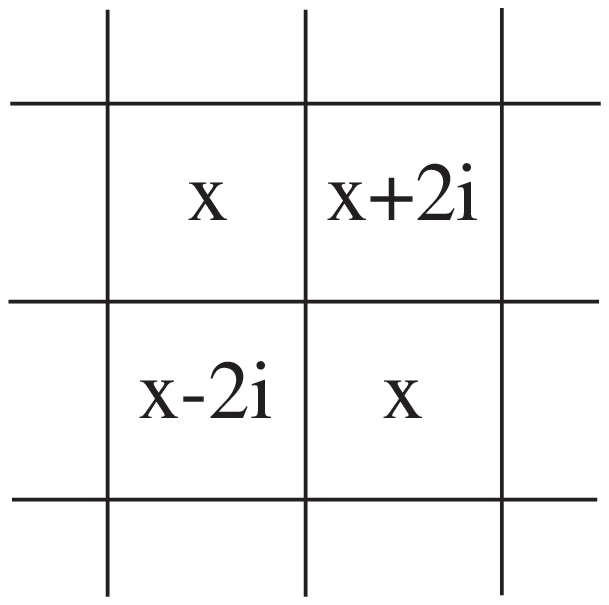}
\caption{Assignment of spectral parameter.}
\label{spec0}
\end{figure}

As in section \ref{survey}, we represent these three terms
 by three boxes with
letters 1,2 and 3, 
\begin{equation}
T_1(u,x)= \framebox[0.4cm][c]{1}_{x} + \framebox[0.4cm][c]{2}_{x} +
\framebox[0.4cm][c]{3}_{x}.  
\label{box-expr-rule}
\end{equation}
One infers from the $sl_2$ example that
a combinatorial aspect may also appear. 
This turns out to be true.
The eigenvalues of fusion QTMs are given by 
the sum of combinations of boxes, 
which can be identified with semi-standard 
Young tableaux  for $sl_3$.
On each diagram, the spectral parameter changes $+2i$ 
from the left to the right and
$-2i$ from the top to the bottom. (Fig \ref{spec0}).

We restrict ourselves to diagrams of the rectangular shape for a while.
Firstly we note that QTM associated to  $2\times m$ ( $3\times m$) 
 Young diagram can be reduced
to the one associated to $1\times m$ (or just scalars). 
This is due to identities,
\begin{eqnarray*}
\begin{tabular}{|l|}
\hline
1 \\
\hline
2 \\
\hline
\end{tabular}
\raise 2mm \vtop{  \hbox{$\phantom{}_{x+i}$} \hbox{$\phantom{}_{x-i}$}  }
&=& \phi(x+\frac{5}{2}i) \phi(x-\frac{5}{2}i) \framebox[0.4cm][c]{$1$}_{x},
\quad
\begin{tabular}{|l|}
\hline
1 \\
\hline
3 \\
\hline
\end{tabular}
\raise 2mm \vtop{  \hbox{$\phantom{}_{x+i}$} \hbox{$\phantom{}_{x-i}$}   }
=\phi(x+\frac{5}{2}i) \phi(x-\frac{5}{2}i) \framebox[0.4cm][c]{$2$}_{x}  \np
\begin{tabular}{|l|}
\hline
2 \\
\hline
3 \\
\hline
\end{tabular}
\raise 2mm \vtop{  \hbox{$\phantom{}_{x+i}$} \hbox{$\phantom{}_{x-i}$}   }
&=&\phi(x+\frac{5}{2}i) \phi(x-\frac{5}{2}i) \framebox[0.4cm][c]{$3$}_{x}
\qquad
\begin{tabular}{|l|l}
\hline
1\\
\hline
2 \\
\hline
3\\
\hline
\end{tabular}
\begin{array}{l}
 \phantom{}_{x+2i}  \\
 \phantom{}_{x}    \\
 \phantom{}_{x-2i}
\end{array}
=\prod_{j=1}^6 \phi(x+(9/2-j)i).
\end{eqnarray*}
Second, eigenvalues of $1\times m$  fusion QTMs have the "duality" in
the following sense.
Let us denote a renormalized $1\times m$  fusion QTM by
$T_m(x)$;
\begin{equation}
T_m(x)= \frac{1}{f_m(x)} 
\sum_{i_1 \le i_2 \le \cdots \le i_m}
\begin{tabular}{|l|l|l|l|}
\hline
$i_1$& $i_2$& $\cdots$& $i_m$ \\
\hline
\end{tabular} .
\label{deft1m}
\end{equation}
The spectral parameters 
are assigned $x-i(m-1) \cdots x+i(m-1)$ from the left to the right. 
The renormalization factor, common factor to all
 expressions of length $m$ tableaux, is given by
$$
f_m(x):= \prod_{j=1}^{m-1} \phi(x\pm i(\frac{2m-1}{2}-j)) .
$$
Hereafter we sometimes denote  $f(x\pm i y) := f(x+iy) f(x-iy)$.

The  resultant $T_m$'s are all degree $2N$ w.r.t. 
$h[x+\hbox{ shift }]$.
Obviously, we have a periodicity due to Boltzmann weights;
\begin{equation}
T_m(x+\frac{10}{3} i) =T_m(x), \qquad (T_m(x+\frac{14}{4} i) =T_m(x))
\label{period}
\end{equation}
for the dilute $A_4$ ($A_6$) model.
From the $sl_3$ structure, together with the above property,
one can prove the following functional relations,
\begin{eqnarray}
T_m(x-i) T_m(x+i) &=&g_m(x) T_m(x) + 
          T_{m+1}(x) T_{m-1}(x),  \quad m \ge 1 \np
g_m(x) &=& \phi(x\pm i(m+3/2)),	 \np
T_{-1}(x) &:=& 0   \np
T_0(x) &:=& f_2(x).
\label{sl3fusion}
\end{eqnarray}

The periodicity for the dilute $A_4$ model , $\phi(x + 10/3i)=\phi(x)$, 
leads to
$g_{m+10}(x)=g_m(x) \,\, m\ge 0$ 
and $g_{7-m}(x)=g_m(x), (0\le m\le 7) $.
From the adjacency matrix, one concludes $T_{8,9}(x)=0$.

Thus functional relations are invariant under the 
transformation, $ T_m(x) \rightarrow T_{7-m}(x) \,\, (m=0,\cdots, 7)$
or $T_{m}(x) \rightarrow T_{m+10}(x) \,\, (m \ge -1)$. 
The symmetry of functional relations is not necessary 
inherited to its solution in general.
We verify, however,  the "duality", 
\begin{equation}
T_m(x)=T_{7-m}(x), \,\, m=0,\cdots, 7
\label{dualonerow}
\end{equation}
and $T_{m+10}(x)=T_{m}(x), m \ge -1$ 
numerically for the largest eigenvalue sector, only which
we are interested in.
These duality relations are also observed and numerically verified for
$A_6$ models with change in the period. See section \ref{summary-A6}.

Functional relations among them, however, do not
possess the desired analytical property, as discussed in \cite{SuzE8}.
We thus introduce other class of QTMs related to skew Young diagrams.

Let $\mu$ and $\lambda$ be a pair of Young diagrams satisfying 
 $\mu_i \ge \lambda_i, \forall i$.
We subtract  a diagram  $\lambda$ from $\mu$. 
We call the result a skew Young diagram $\mu-\lambda$, consisted of
$(\mu_1-\lambda_1,\mu_2-\lambda_2,\cdots )$ boxes.
\begin{figure}[hbtp]
\centering
  \includegraphics[width=3cm]{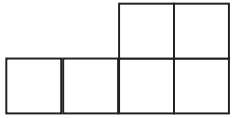}
\caption{An example of a skew Young Table, (4,4)-(2)}
\label{skew1}
\end{figure}
In the theory of symmetric polynomials,
the Jacobi-Trudi formula tells that a complex Schur function
associated to  a skew Young diagram can be
expressed by a determinant of a matrix, of which elements are
given by elementary Schur functions  associate to "one-row" diagrams
or " one-column" ones.
The quite parallel formula holds for the present situation, 
 which we call the "quantum" Jacobi-Trudi formula  \cite{BR, KS, KOS}.

Consider a set of semi-standard skew Young tableaux of the shape $\mu-\lambda$.
We assign  an expression to each table.
The spectral parameter of  the "top-left" box is fixed to 
$x+i(\mu_1'-\mu_1)$ 
\footnote{
Using this opportunity we remark misprints in the 4th row in the second
paragraph of section 5 and the caption of Fig2 in \cite{SuzE8}.}
where $\mu'_1$ denotes the depth of the tableaux.
\begin{figure}[hbtp]
\centering
  \includegraphics[width=4cm]{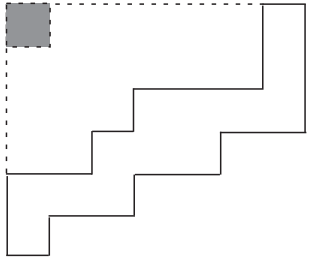}
\caption{The spectral parameter $x+i(\mu_1'-\mu_1)$ is  assigned to the hatched
place.}
\label{spectral}
\end{figure}
One identifies each box in a table with an expression 
under the rule (\ref{box-expr-rule}) with
 the shift of the spectral parameter.
Then the product over all constituting boxes yields
the desired expression for the table.
\begin{thm}
Let ${\cal T}_{\mu/\lambda}(x)$ be the sum 
over the resultant expressions divided by a common factor,
$
\prod_{j=1}^{\mu'_1} f_{\mu_j-\lambda_j} 
  (x+i(\mu_1' -\mu_1+\mu_j+\lambda_j-2j+1)).
$
Then the following  equality holds,
\begin{equation}
{\cal T}_{\mu/\lambda}(x) =
  \hbox{det }_{1\le j,k\le \mu_1'} 
         ( T_{\mu_j-\lambda_k-j+k} 
            (x+i(\mu_1' -\mu_1+\mu_j+\lambda_k-j-k+1)) )
\label{qJT}
\end{equation}
where $T_{m<0}:=0$.
\end{thm}

The proof is quite similar to the one for Young tableaux\cite{KS, KOS}.
One must only keep in mind that the allowed position of
a box is restricted by its spectral parameter.

The crucial fact is that the
so defined  ${\cal T}_{\mu/\lambda}(x)$ is an analytic function of $x$
due to BAE, and contains  $T_1(x)$ as a special case.
The former is not so obvious from the original definition by the tableaux, but
it follows trivially from the quantum Jacobi-Trudi formula.

In the same spirit, we introduce  $\Lambda_{\mu/\lambda}(x)$, 
which is analytic under BAE,
 by putting $T_{m\ge 8}(x)=0$ in  ${\cal T}_{\mu/\lambda}(x)$,
\begin{equation}
 \Lambda_{\mu/\lambda}(x):={\cal T}_{\mu/\lambda}(x) /.\{ T_{m\ge 8}(x)=0 \},
\label{qJTLambda}
\end{equation}
for the dilute $A_4$ model.
The pole-free property of $\Lambda_{\mu/\lambda}(x)$ is obvious from 
(\ref{qJT}).
For the dilute $A_6$ model, we define $\Lambda_{\mu/\lambda}(x)$
by  setting $T_{m\ge 11}(x)=0$.

%
%
\section{Kink Transfer Matrix}\label{kinkQTM}
In following few sections, we restrict our discussion to
the dilute $A_4$ model. 
We will summarize results for the dilute $A_6$ model
in section \ref{summary-A6}.

For the $E_8$ case, the vector representation is minimal and other
representations are constructed by fusion of it.
Eigenvalues of associated transfer matrices of the dilute $A_3$ model
are thus 
derived from products of $T_1(x+{\hbox{some shift}})$.

This is no longer true for the dilute $A_4$ model.
Let $W_a(x)$ be the Yangian highest weight module associated  to
the node $a$ in Figure \ref{dynkin}.
Through several evidences, we identify $T_1(x)$ of the model 
with the transfer matrix connected to $W_1(x)$, which is not minimal.
The most fundamental
object in $E_7$ is $W_6(x)$ rather than $W_1(x)$.
Any other object may be constructed from $T^{(6)}(x)$,
the transfer matrix of $W_6(x)$,  of which explicit form is not
known from the dilute $A_4$ model.
The determination of explicit form of $T^{(6)}(x)$ is thus vital in 
the present approach.
For this purpose, homomorphisms of $U_q(\hat{\mathfrak{g}})$ modules 
deserve attentions.
They lead to non-trivial
algebraic relations among  $T^{(6)}(x)$ and other QTMs.
Although such information is not available for $U_q(\hat{E_7})$, there 
exists a list of homomorphisms of  $Y(E_7)$ modules in \cite{Nakanishi},

\begin{eqnarray}
W_{6-a}(x) &\hookrightarrow& W_{7-a} (x-\frac{1}{6}i) \otimes 
 W_{6} (x+\frac{a}{6}i), \qquad a=1,2,3  
 \label{homoone} \\
W_1(x) &\hookrightarrow& W_6(x+\frac{5}{6}i) \otimes W_6(x-\frac{5}{6}i) 
 \label{homotwo}\\
C &\hookrightarrow& W_6(x+\frac{3}{2}i) \otimes W_6(x-\frac{3}{2}i) 
 \label{homothree}\\
W_2(x) &\hookrightarrow& W_1(x+\frac{1}{6}i) \otimes W_1(x-\frac{1}{6}i)
 \label{homofour} \\
W_7(x) &\hookrightarrow& W_1(x+\frac{1}{2}i) \otimes W_6(x-\frac{2}{3}i).
 \label{homofive} 
\end{eqnarray}
Note that the normalization of spectral parameter is
different from \cite{Nakanishi}.

For example, the second relation implies,
$$
T^{(6)}(x+i\frac{5}{6}) T^{(6)}(x-i\frac{5}{6})=T^{(1)}(x)  + \cdots .
$$
After trials and errors, we find the following ansatz 
 compatible with the above homomorphisms ,
\begin{eqnarray}
T^{(6)}(x)&=& \frac{1}{\sqrt{2}} \Bigl (
 w^2 \phi(x+\frac{2}{3}i) \frac{Q(x+\frac{2}{3}i)}{Q(x-\frac{1}{3}i)}+
 w \phi(x-\frac{4}{3}i) \frac{Q(x-\frac{4}{3}i) Q(x+\frac{5}{3}i)}
                        { Q(x-\frac{1}{3}i) Q(x+i)}  \np
 & &+ \phi(x) \frac{Q(x+\frac{5}{3}i) Q(x)}
                        { Q(x+i) Q(x-i)} +		
 \frac{1}{w} \phi(x+\frac{4}{3}i) \frac{Q(x+\frac{4}{3}i) Q(x-\frac{5}{3}i)}
                        { Q(x+\frac{1}{3}i) Q(x-i)} \np
 & &+\frac{1}{w^2} \phi(x-\frac{2}{3}i) \frac{Q(x-\frac{2}{3}i)}{Q(x+\frac{1}{3}i)}
  \Bigr ).  \label{t6ansatz}
\end{eqnarray}

As explicit RSOS weights are not yet derived, it may  be inappropriate
to call $T^{(6)}(x)$ as the eigenvalue of transfer matrix.
In the following discussion, however, we do not use the assumption
that it coincides with the actual eigenvalue of transfer matrix of
$W_6$.
Rather, we simply use facts (1) it is pole free under BAE
(2) it satisfies the desired relations expected from eqs.
(\ref{homotwo}) and  (\ref{homothree}).  
Readers should understand this terminology just as a "nickname".

The following functional relations between $T^{(6)}(x)$ 
and $T_m(x)$ will facilitate discussions in later sections.
\begin{lem}
\begin{eqnarray}
T^{(6)}(x+\dl i) T^{(6)}(x-\dl i)&=&T_2(x+\frac{10}{6}i)+T_0(x) \label{tstsone}\\
T^{(6)}(x+\frac{3}{6}i) T^{(6)}(x-\frac{3}{6}i) &=&T_3(x) \label{tststwo} \\
T^{(6)}(x+\frac{5}{6}i) T^{(6)}(x-\frac{5}{6}i) 
  &=&T_1(x)+T_1(x+\frac{10}{6}i) \label{tststhree} \\
T^{(6)}(x+\frac{7}{6}i) T^{(6)}(x-\frac{7}{6}i) &=&T_3(x+\frac{10}{6}i) 
\label{tstsfour} \\
T^{(6)}(x+\frac{9}{6}i) T^{(6)}(x-\frac{9}{6}i)&=&
T_2(x)+T_0(x+\frac{10}{6}i). \label{tstsfive}
\end{eqnarray}
\end{lem}
%
\noindent{Proof: } 
The duality relation (\ref{dualonerow}) plays a fundamental role in the proof.
The direct substitutions of eq.(\ref{t6ansatz}) in the lhs of
eqs. (\ref{tststwo}), (\ref{tststhree})  and (\ref{tstsfive}) 
yield  $1/2( T_3(x)+T_4(x))$ , 
$1/2( T_1(x+i\frac{10}{6})+T_6(x+\frac{10}{6}i)+2 T_1(x))$  and
$1/2( T_2(x)+T_5(x))+ T_0(x+\frac{10}{6}i) $, respectively. 
We remark that non-trivial cancellation of terms occurs
due to $w^5 = -1$.  \pn
These results coincide with the rhs  due to dualities
 $T_3 =T_4$,  $T_1 =T_6$ and  $T_2 =T_5$. \pn
Rests follow from these by $x \rightarrow x+i\frac{10}{6}$. 
$\Fsquare(0.2cm,  )$

These relations suggest the underlying homomorphisms between 
$U_q(\hat{E_7})$ and $U_{q'}(A^{(2)}_2)$ modules at $q=\exp(i \pi/20),
q'=\exp(i 3 \pi/10)$.
This may be an interesting but an independent subject from the present problem
thus we will not go into detail here.
%
%
\section{Fusion Quantum transfer matrices and the $T-$ system
for the dilute $A_4$ model }\label{Tsys-A4}
Having defined the "kink" QTM $T^{(6)}$, we are in position to introduce
other QTMs and explore functional relations among them.

First we present QTMs defined by skew Young tableaux.
\begin{df}
\begin{eqnarray}
T^{(1)}(x) &=& \Lambda_{(1)}(x) \quad (=T_1(x))  \label{defTone} \\
T^{(2)}(x) &=&\frac{1}{\phi(x-\frac{5}{3}i)}
     \Lambda_{(6,1)}(x-\frac{5}{6}i)  \label{defTtwo} \\
T^{(3)}(x) &=&\frac{1}{\phi(x\pm \frac{3}{2}i)}
\Lambda_{(11,6,6)/(5,5)}(x)  \label{defTthree} \\
T^{(5)}(x) &=& \Lambda_{(2)}(x+\frac{5}{3}i)  \quad 
(=T_2(x+\frac{5}{3}i) ). \label{defTfive} 
\end{eqnarray}
\end{df}
We have two comments. \pn
First, one can  equivalently rewrite (\ref{defTtwo}) and (\ref{defTthree})
in terms of  $\Lambda_{(6,6)/(5)}$ or $\Lambda_{(6,6,1)/(5)}(x)$
using relations,
\begin{eqnarray}
\Lambda_{(6,6)/(5)}(x) &=& \Lambda_{(6,1)}(x+5i)  \label{Lambdadual1} \\ 
\Lambda_{(6,6,1)/(5)}(x)&=&\Lambda_{(11,6,6)/(5,5)}(x).  \label{Lambdadual2}
\end{eqnarray}
These are outcome of the quantum Jacobi-Trudi 
formula and duality (\ref{dualonerow}).  \pn
The second comment concerns the complex conjugate property.
In the largest eigenvalue sector of QTM, we confirm numerically
$Q(x)=\overline{Q(\bar{x})}$.
The explicit forms of DVFs in eq.(\ref{t1expr}) and eq.(\ref{t6ansatz})
then conclude 
$T^{(1)}(x) =\overline{ T^{(1)}(\bar{x}) },  
T^{(6)}(x) =\overline{ T^{(6)}(\bar{x}) }$.
By remembering $5/3i$ equals to the half period of our elliptic function
$\theta_1(x)$, one also verifies  $T^{(5)}(x)=\overline{ T^{(5)}(\bar{x}) }$. 
The  conjugate  property of $T^{(2)}(x) $
and $T^{(3)}(x)$ is less obvious. 
One can nevertheless show this by the use of
 the quantum Jacobi-Trudi  formula.

$T^{(6)}$ comes into expressions of remaining QTMs. 
The Yangian homomorphisms turn out to be useful in deriving their explicit forms.
We take $T^{(4)}(x)$, related to $W_4(x)$,  for instance.
As argued in section \ref{ABA}, we identify  an analytic set  with 
an affine irreducible module.
Thus  eq. (\ref{homoone}) for $a=2$ implies,
$$
T^{(5)}(x-i/6) T^{(6)}(x+i/3) \sim T^{(4)}(x) +T'(x).
$$
That is, the product of DVFs  $T^{(5)}(x-I/6) T^{(6)}(x+i/3)$ 
decomposes into two (or more)
subsets and each subset is analytic within itself.
We look at the explicit DVF of the lhs and find that it contains analytic subset
given by  
$\phi(x+\frac{4}{3}i)\phi(x-i) T^{(6)}(x-\frac{1}{3}i)$.
The sum of remaining terms must be analytic under BAE.
We identify them as $ T^{(4)}(x)$.
In a similar way, we deduce $T^{(7)}(x)$.

\begin{df}
\begin{eqnarray}
T^{(4)}(x) &=& 
(T^{(5)}(x-\frac{1}{6}i) T^{(6)}(x+\frac{1}{3}i)-
\phi(x+\frac{2}{3}i)\phi(x-\frac{1}{3}i) T^{(6)}(x-\frac{1}{3}i))   \np
 \label{defTfour} \\
T^{(7)}(x) &=& \frac{1}{ \phi(x-\frac{4}{3}i)}
(T^{(1)}(x+\frac{1}{2}i) T^{(6)}(x-\frac{2}{3}i)-
\phi(x)\phi(x-i) T^{(6)}(x+\frac{4 }{3}i)). \np
\label{defTseven} 
\end{eqnarray}
\end{df}

\noindent The common factor $\frac{1}{ \phi(x-\frac{4}{3}i)}$
is divided out for $T^{(7)}(x)$.
We note that the conjugate property, discussed for other $T^{(a)}$s, 
is also verified for $T^{(7)}(x)$ when it is
written in terms of its explicit DVF.
On the other hand, the property is  not so apparent for $T^{(4)}(x)$,
although one can prove it by a different route. 
See the discussion after 
the proof of eq.(\ref{t5}).

We are now ready to describe the statement as to functional relations among
QTMs .

\begin{prop}
The above QTMs enjoy the following $T-$system, 
\begin{eqnarray}
T^{(1)}(x-\dl i) T^{(1)}(x+\dl i) &=& \phi(x-\frac{5}{3}i) T^{(2)}(x)
               +T_0(x\pm \frac{5}{6}i) , \label{t1}\\
T^{(2)}(x-\dl i) T^{(2)}(x+\dl i) &=& T_0(x) T_0(x\pm \frac{4}{6}i)+
 T^{(1)}(x) T^{(3)}(x) ,    \label{t2}\\
T^{(3)}(x-\dl i) T^{(3)}(x+\dl i) &=&
T_0(x\pm \frac{1}{2} i)    T_0(x\pm \frac{1}{6}i)
+T^{(2)}(x) T^{(4)}(x) T^{(7)}(x),   \np
\label{t3}\\
T^{(4)}(x-\dl i) T^{(4)}(x+\dl i) &=&
 T_0(x) T_0(x\pm \frac{1}{3}i) +
 T^{(3)}(x) T^{(5)}(x),    \label{t4} \\
T^{(5)}(x-\dl i) T^{(5)}(x+\dl i) &=&
T_0(x \pm \dl i) +
T^{(4)}(x) T^{(6)}(x) ,  \label{t5}\\
T^{(6)}(x-\dl i) T^{(6)}(x+\dl i) &=&
T_0(x ) 
+ T^{(5)}(x) ,  \label{t6}\\
T^{(7)}(x-\dl i) T^{(7)}(x+\dl i) &=&
T_0(x\pm\frac{1}{3}i) + T^{(3)}(x). \label{t7} 
\end{eqnarray}
\end{prop}

These coincide with the $T-$ system for $E_7$ proposed in
\cite{KNS1} in a different context.

Note that eq.(\ref{t6}) has already been proven in eq.(\ref{tstsone}) by
 the definition (\ref{defTfive}).

\noindent Proof of eqs.(\ref{t1}) and (\ref{t2}).  \par
\begin{figure}[hbtp]
\centering
  \includegraphics[width=12cm]{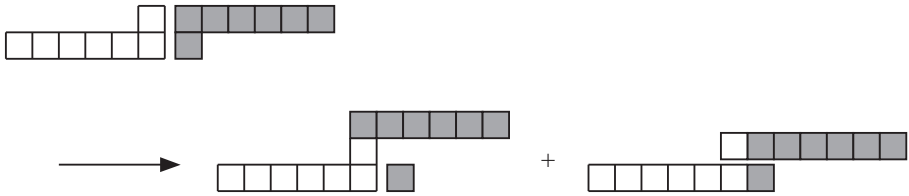}
\caption{ The decomposition of
$\Lambda_{(6,6)/(5)}(x) \Lambda_{(6,1)}(x+12 i)$.}
\label{T2T2prod}
\end{figure}

\noindent Take the simpler case (\ref{t1}) first. 
We consider the decomposition of
the product $T_1(x-6i) T_6(x+i)$. 
Quite similar to combinatorics
of semi-standard tableaux, we have
$$
T_1(x-6i) T_6(x+i) = \Lambda_{(6,1)}(x)+T_7(x)T_0(x-5i),
$$
which can also be verified from the quantum Jacobi-Trudi formula.
By utilizing dualities $T_6(x)=T_1(x)$, $T_7(x)=T_0(x)$ and the periodicity
, one recovers eq.(\ref{t1}) after the shift
 $x \rightarrow x-\frac{5}{6} i $ in both sides.\pn
Eq. (\ref{t2}) also follows by considering the decomposition of
$\Lambda_{(6,6)/(5)}(x) \Lambda_{(6,1)}(x+12 i)$. 
See Fig. \ref{T2T2prod}.
The lhs of the arrow contains  two Young diagrams corresponding to
$\Lambda_{(6,6)/(5)}(x)$ and $\Lambda_{(6,1)}(x+12 i)$ (hatched) 
in the relative position, 
compatible with the shift in the spectral parameter.
We then employ  the "recombination" of
boxes, as in usual Young diagrams. 
Two rules need particular attention.
First, we must put box so as to match the spectral parameter.
Second, due to the condition $T_{m\ge 8}(x)=0$ in the definition 
of $\Lambda_{\mu/\lambda}(x)$, the width of a column in the 
resultant  Young diagram must be less equal to 7.
Then two terms result from the "recombination"  as 
depicted in the rhs of the arrow,
$\Lambda_{(11,6,6)/(5,5)}(x-8i) T_1(x+6i) $ and $ T_7(x) T_7(x+12i)$.
We still leave hatch to boxes which once belong to $\Lambda_{(6,1)}(x+12 i)$.
Note that the figure represents the  equality of the DVFs
in terms of boxes.
Thus it needs proper normalization factors, viewed as
the relation between $\Lambda_{\mu/\lambda}(x)$ or $T_m(x)$, 
due to scalar factors in the definitions (\ref{deft1m})
of $T_m(x)$   and of ${\cal T}_{\mu/\lambda}(x)$
(and then $\Lambda_{\mu/\lambda}(x)$) in Theorem 1.
Then one finds, with the property  (\ref{Lambdadual1}), 
\begin{eqnarray*} 
& & \Lambda_{(6,1)}(x-5i)  \Lambda_{(6,1)}(x+12i) =
    \Lambda_{(11,6,6)/(5,5)}(x-8i) T_1(x+6i) \\
& &  +  \phi(x-\frac{9}{2}i) \phi(x-\frac{11}{2}i)  
 \phi(x+\frac{13}{2}i) \phi(x+\frac{15}{2}i) T_7(x) T_7(x+12 i) 
\end{eqnarray*} 
where the factor in front of  $T_7(x) T_7(x+12 i) $ comes from
 $f_7(x) f_7(x+12 i) /f_6(x+i) f_6(x+13i)$.
Finally let us shift the spectral parameter  by $+i4/6$
 and use the periodicity (\ref{period}).
Then eq.(\ref{t2}) follows from
definitions (\ref{defTtwo}),(\ref{defTthree}) and
$T_7=T_0$.

$\fsquare(0.2cm, )$
%

\noindent Proof of eq.(\ref{t5}). \par
The proof utilizes
(\ref{tstsone}) as follows. 
Consider the product $T^{(4)}(x) T^{(6)}(x)$. 
Substituting eq. (\ref{defTfour}),  we have
$$
T^{(4)}(x) T^{(6)}(x)= T^{(5)}(x-\dl i) T^{(6)}(x) T^{(6)}(x+\frac{1}{3}i)-
T_0(x+\dl i)  T^{(6)}(x) T^{(6)}(x-\frac{1}{3}i).
$$
One applies the rule (\ref{tstsone}) to the product of two  $T^{(6)}$s 
by shifting $x \pm i/6$. The result leads to 
$$
T^{(4)}(x) T^{(6)}(x)= T^{(5)}(x-\dl i)T^{(5)}(x+\dl i)-T_0(x \pm \dl i) ,
$$
which coincides with eq.(\ref{t5}).$\fsquare(0.2cm, )$

As mentioned previously,  the complex conjugate property of $T^{(4)}(x)$
is not obvious from its DVF.
Instead, we can now shown it from the established relation (\ref{t5}) with the help 
of the conjugate property of $T^{(5)}(x)$ and $T^{(6)}(x)$ commented below
(\ref{defTfive}).


To prove remaining relations, we need to prepare further lemmas.

\begin{lem}
The following decompositions are valid ,
\begin{eqnarray}
& &T^{(5)}(x+\frac{1}{3}i) T^{(5)}(x-\frac{1}{3}i) =
T_0(x) T_4(x) + T^{(3)}(x)   \label{t5t51} \\
& &T^{(5)}(x+\frac{1}{3}i) T^{(5)}(x-\frac{1}{3}i) =
 T_4(x)(T_0(x)+T^{(5)}(x) ) - T_0(x\pm \frac{1}{3}i) \np
& & -(T_0(x+ \frac{1}{3}i) T^{(5)}(x- \frac{1}{3}i) +
      T_0(x- \frac{1}{3}i) T^{(5)}(x+\frac{1}{3}i)).  \label{t5t52} 
\end{eqnarray}
\label{t5t5}
\end{lem}
%
\noindent Proof of Lemma \ref{t5t5}. 
The relation (\ref{t5t51}) is checked by comparing DVFs of both sides
directly.
To prove eq.(\ref{t5t52}), we rewrite $T^{(5)}$ in the lhs in terms of
$T^{(6)}$ by using eq.(\ref{tstsone}),
\begin{eqnarray*}
& &T^{(5)}(x+\frac{1}{3}i) T^{(5)}(x-\frac{1}{3}i) =
T_0(x\pm \frac{1}{3}i) \\
& &+T^{(6)}(x-\frac{1}{6}i) T^{(6)}(x+\frac{1}{6}i)
T^{(6)}(x-\frac{1}{2}i) T^{(6)}(x+\frac{1}{2}i)\\
& &-(T_0(x+\frac{1}{3}i)T^{(6)}(x-\frac{1}{6}i)T^{(6)}(x-\frac{1}{2}i)+
 T_0(x-\frac{1}{3}i)T^{(6)}(x+\frac{1}{6}i)T^{(6)}(x+\frac{1}{2}i))
\end{eqnarray*}
By applying eqs.(\ref{tstsone}) and (\ref{tststwo}), one reaches the rhs of 
(\ref{t5t52} ). $\fsquare(0.2cm, )$

For later use, we rearrange the sum of eqs.(\ref{t5t51}) and (\ref{t5t52}),
\begin{eqnarray}
& & \Bigl ( 
       2 T_0(x) T_4(x) + T^{(3)}(x)+T_4(x) T^{(5)}(x) 
       -T_0(x+\frac{1}{3}i)T^{(5)}(x-\frac{1}{3}i)    \np
& &  -T_0(x-\frac{1}{3}i)T^{(5)}(x+\frac{1}{3}i) 
  -2 T^{(5)}(x-\frac{1}{3}i) T^{(5)}(x+\frac{1}{3}i) \Bigr ) = 
  T_0(x\pm \frac{1}{3}i). \np
\label{t5t5res}
\end{eqnarray}


\begin{lem}
\begin{eqnarray*}
& &T_0(x\pm i) T_3(x) + T_1(x-2i)T_1(x) T_1(x+2i)  \\
& &- T_0(x-i)T_1(x+2 i) T_2(x-i) -T_0(x+i)T_1(x-2 i) T_2(x+i) \\
& &= \phi(x\pm \frac{7}{2}i)\phi(x\pm \frac{5}{2}i)\phi(x\pm \frac{3}{2}i).
\end{eqnarray*}
\label{triple}
\end{lem}

\noindent
This is the analogue of the relation,
$$
\raise 2ex
\hbox{$
     \fsquare(0.3cm, ) \fsquare(0.3cm, ) \fsquare(0.3cm, )
        +
      \fsquare(0.3cm, )\otimes \fsquare(0.3cm, ) \otimes \fsquare(0.3cm, )
        -
      \fsquare(0.3cm, )\otimes \fsquare(0.3cm, ) \fsquare(0.3cm, )
        -\fsquare(0.3cm, )\fsquare(0.3cm,  )\otimes  \fsquare(0.3cm, )
	=$}
\triple(0.3cm),
$$
and it can be shown in a similar manner.

In proving eq.(\ref{t3}), one needs decomposition of a huge QTM.
\begin{lem}
\begin{eqnarray*}
& &\Lambda_{(11,11,6,6)/(10,5,5)}(x) =
\phi(x+\frac{1}{2}i)\phi(x-\frac{5}{2}i) \times \\
& & 
\Bigl (
   T_1(x+3i)T_2(x+i)T_2(x-3i) -T_0(x+2i)T_1(x-2i) T_2(x+i) \\
& & -T_0(x-i)T_1(x+3i)T_3(x-i) + T_0(x-i)T_0(x+2i)T_4(x) \Bigr ).
\end{eqnarray*}
\label{huge}
\end{lem}

\noindent Proof of Lemma \ref{huge}. 
For convenience, we go back to the original definition of
${\cal T}_{(11,11,6,6)/(10,5,5)}(x)$.
Thanks to the semi-standard condition, it decomposed into pieces.
\begin{figure}[hbtp]
\centering
  \includegraphics[width=8cm]{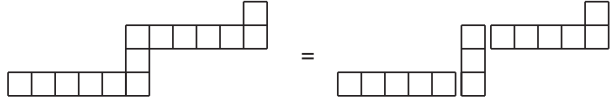}
\caption{Decomposition of the skew diagram (11,11,6,6)/(10,5,5).
The height 3 piece in the middle of the rhs reduces to a known scalar factor.}
\label{hugedecomp}
\end{figure}
We consequently have 
\begin{eqnarray}
& &f_6(x+6i) f_6(x-8i) {\cal T}_{(11,11,6,6)/(10,5,5)}(x) =
\frac{f_5(x-i)f_5(x-9i)f_5(x+7i)}{f_2(x-i)} \np
& &\times T_5(x-9i)  
 \Bigl (
 T_1(x+13 i)T_5(x+5i) -T_0(x+12i) T_6(x+8i) \Bigl ).
\label{huge1} \np
\end{eqnarray}
On the other hand, the quantum Jacobi-Trudi formula relates the same quantity to
\begin{eqnarray}
& &f_6(x+6i) f_6(x-8i) \Lambda_{(11,11,6,6)/(10,5,5)}(x) \np
& &+ f_{13}(x-i)T_1(x+13i) T_{13}(x-i) -f_{14}(x)T_{14}(x). 
\label{huge2} \np
\end{eqnarray} 
Then the equality (\ref{huge1})=(\ref{huge2}) with properties 
$T_{m+10}(x) = T_m(x)$ and $T_{7-m}(x)=T_m(x)$   leads to
Lemma \ref{huge} after renormalization. $\fsquare(0.2cm, )$
%

Our final lemma concerns the decomposition of $T^{(4)}(x) T^{(7)}(x)$.
\begin{lem}
\begin{eqnarray*}
& & \phi(x-\frac{4}{3}i) T^{(4)}(x) T^{(7)}(x)=\\
& & T_2(x+\frac{11}{6}i)
\Bigl \{
   T_1(x+\frac{1}{2}i)T_2(x+\frac{7}{6}i)-T_0(x+\frac{17}{6}i) T_1(x-\frac{7}{6}i)  
   \Bigr \}  \\
& & -T_0(x-\frac{1}{6}i)T_1(x+\frac{1}{2}i)T_3(x-\frac{1}{6}i) +
    T_0(x-\frac{1}{6}i)T_0(x-\frac{1}{2}i)T_3(x+\frac{5}{6}i) .
\end{eqnarray*}
\label{t4t7}
\end{lem}
%
\noindent{Proof of Lemma \ref{t4t7}. } 
We use a trick; 
the complex conjugate property, established soon
below the proof of eq.(\ref{t5}), allows us to replace  $T^{(4)}(x)$ by  
$\overline{T^{(4)}(x)}$ in the lhs.
After this, we substitute definitions (\ref{defTfour}) and 
(\ref{defTseven}) into the lhs,
\begin{eqnarray*}
& &\phi(x-\frac{4}{3}i) \overline{T^{(4)}(x)} T^{(7)}(x)=\\
& & T_2(x+\frac{11}{6}i) T_1(x+\frac{1}{2}i) T^{(6)}(x-\frac{1}{2}i+\dl i)
    T^{(6)}(x-\frac{1}{2}i-\dl i) \\
& &-\phi(x)\phi(x-i) T_2(x+\frac{11}{6}i) T^{(6)}(x+\frac{1}{2}i+\frac{5}{6}i)
    T^{(6)}(x+\frac{1}{2}i-\frac{5}{6}i ) \\
& & -\phi(x+\frac{1}{3}i)\phi(x-\frac{2}{3}i) T_1(x+\frac{1}{2}i)
    T^{(6)}(x-\frac{1}{6}i+\frac{1}{2}i) T^{(6)}(x-\frac{1}{6}i-\frac{1}{2}i )\\
& &+\phi(x-i)\phi(x-\frac{2}{3}i)\phi(x)\phi(x+\frac{1}{3}i)
    T^{(6)}(x+\frac{5i}{6}+\frac{1}{2}i) T^{(6)}(x+\frac{5i}{6}-\frac{1}{2}i).
\end{eqnarray*}
where we have used definitions (\ref{defTone}) and (\ref{defTfive}) 
to rewrite $T^{(1)}$ and $T^{(5)}$ by $T_1$ and $T_2$.
Thanks to the trick, the differences in arguments of products of
$T^{(6)}$ are such that one can apply 
eqs. (\ref{tstsone}), (\ref{tststwo}) and (\ref{tststhree}).
The result of the application then
 agrees with the rhs of the equality in Lemma \ref{t4t7}.
$\fsquare(0.2cm, )$

By  comparing the rhs of Lemma \ref{huge} and Lemma \ref{t4t7}
with use of the duality $T_3=T_4$,
one notices the following equality.

\begin{lem}
$$
\Lambda_{(11,11,6,6)/(10,5,5)}(x+\frac{5}{6}i) =
\phi(x \pm \frac{8}{6}i) \phi(x- \frac{10}{6}i) T^{(4)}(x) T^{(7)}(x). 
$$
\label{huget4t7}
\end{lem}


With these preparations, we prove remaining relations.

\noindent{Proof of (\ref{t4}). }

Let us rewrite the lhs by  substituting the definition (\ref{defTfour}).
In doing so, we employ a similar trick to the above; 
we replace $T^{(4)}(x+\dl i)$ in the lhs by 
$\overline{T^{(4)}(x-\dl i)}$. 
Then it follows,  
\begin{eqnarray*}
& & T^{(4)}(x+\dl i) T^{(4)}(x-\dl i)=\\
& & T^{(5)}(x-\frac{1}{3}i) T^{(5)}(x+\frac{1}{3}i)
     T^{(6)}(x-\dl i) T^{(6)}(x+\dl i)    \\
& &-T_0(x)T^{(5)}(x-\frac{1}{3}i) 
       T^{(6)}(x+\frac{1}{3}i-\dl i) T^{(6)}(x+\frac{1}{3}i+\dl i)
 \\
& &-T_0(x)T^{(5)}(x+\frac{1}{3}i) 
       T^{(6)}(x-\frac{1}{3}i-\dl i) T^{(6)}(x-\frac{1}{3}i+\dl i) \\
& &+ (T_0(x))^2 
    T^{(6)}(x-\frac{1}{2}i) T^{(6)}(x+\frac{1}{2}i).
\end{eqnarray*}
In this form, eq. (\ref{tstsone}) applies to the first three terms 
of the rhs,  and  eq.(\ref{tststwo}) can be used in the fourth.
Also we use eq. (\ref{t5t51}) 
the first term of the rhs.
The result reads, 
\begin{eqnarray*}
& & T^{(4)}(x+\dl i) T^{(4)}(x-\dl i)= 
    T^{(3)}(x) T^{(5)}(x)    \\
& &+   T_0(x) \Bigl ( 
       2 T_0(x) T_4(x) + T^{(3)}(x)+T_4(x) T^{(5)}(x) \\
&  &  -T_0(x+\frac{1}{3}i)T^{(5)}(x-\frac{1}{3}i)
      -T_0(x-\frac{1}{3}i)T^{(5)}(x+\frac{1}{3}i)  \\
&  &  -2 T^{(5)}(x-\frac{1}{3}i) T^{(5)}(x+\frac{1}{3}i) \Bigr ),
\end{eqnarray*}
where we represent $T_2(x)$ by $T^{(5)}(x+5/3 i)$ and use the duality $T_3=T_4$.
One notices that the content of bracket in the rhs
reduces to $T_0(x\pm \frac{1}{3}i)$ due to eq.(\ref{t5t5res}), 
which completes the proof. $\fsquare(0.2cm , )$
%

\noindent{Proof of (\ref{t7}). }
In the same manner, 
we start from the equivalent expression 
$\overline{T^{(7)}(x-\dl i)} T^{(7)}(x-\dl i)$ and substitute 
(\ref{defTseven}).
After rearrangement using eqs. (\ref{tststhree}),(\ref{tstsfour}) and
(\ref{tstsfive}), we find this expression  equal to
\begin{eqnarray*}
& & T^{(7)}(x+\dl i) T^{(7)}(x-\dl i)= 
      \frac{1}{\phi(x\pm\frac{3}{2}i)}
   \Bigl ( 
        T_0(x\pm\frac{2}{3}i) T_3(x+\frac{10}{6}i)    \\
&  &    + T_1(x-\frac{1}{3}i) T_1(x+\frac{1}{3}i) T_1(x+\frac{10}{6}i)
        + T_1(x-\frac{1}{3}i) T_1(x+\frac{1}{3}i) T_1(x)   \\
&  &  -T_0(x+\frac{2}{3}i)T_0(x+i)T_1(x+\frac{1}{3}i) 
      -T_0(x+\frac{2}{3}i)T_1(x+\frac{1}{3}i) T_2(x+\frac{2}{3}i) \\
&  &   -T_0(x-\frac{2}{3}i)T_0(x-i)T_1(x-\frac{1}{3}i) 
      -T_0(x-\frac{2}{3}i)T_1(x-\frac{1}{3}i) T_2(x-\frac{2}{3}i) 
	  	     \Bigr ).
\end{eqnarray*}
Let us subtract $ T^{(3)}(x)$ from the above.
Note that  $ T^{(3)}(x)$ should be understood as the result of
the application of (\ref{qJTLambda}) to 
(\ref{defTthree}) and of the duality $T_7 \rightarrow T_0$.
Then we have 
\begin{eqnarray*}
& & T^{(7)}(x+\dl i) T^{(7)}(x-\dl i)-  T^{(3)}(x)  = \\
& &  \frac{1}{\phi(x\pm\frac{3}{2}i)}
  \Bigl ( 
        T_0(x\pm\frac{2}{3}i) T_3(x+\frac{10}{6}i) +
	     T_1(x-\frac{1}{3}i) T_1(x+\frac{1}{3}i) T_1(x+\frac{10}{6}i)  \\
&  &  -T_0(x+\frac{2}{3}i)T_1(x+\frac{1}{3}i) T_2(x+\frac{2}{3}i) 
     -T_0(x-\frac{2}{3}i)T_1(x-\frac{1}{3}i) T_2(x-\frac{2}{3}i) 
  	     \Bigr ).
\end{eqnarray*}
The content of the bracket is
identical to the lhs of Lemma \ref{triple}
with the shift $x \rightarrow x+10/6 i$ by noticing
the periodicity (\ref{period}).
Immediately, one verifies that the rhs reduces to $T_0(x\pm 1/3 i)$ , and
( \ref{t7}) is proven. $\fsquare(0.2cm, )$
%

\begin{figure}[hbtp]
\centering
  \includegraphics[width=9cm]{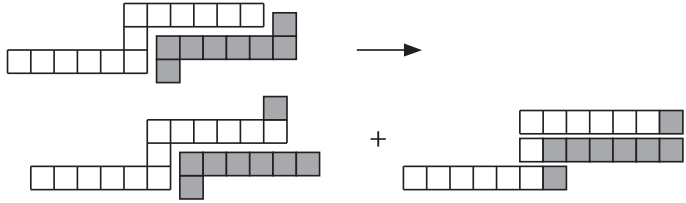}
\caption{Graphical rule for decomposition of two diagram 
(11,6,6)/(5,5) and (6,6,1)/(5).}
\label{T3T3prod}
\end{figure}

\noindent{Proof of ( \ref{t3}). }
The decomposition of $\Lambda_{(11,6,6)/(5,5)}(x) \Lambda_{(6,6,1)/(5)}(x+7i)$
can be done  by the formula (\ref{qJTLambda}).
Equivalently, we can argue it in a graphic manner as shown in Fig. \ref{T3T3prod}, 
just as in Fig. \ref{T2T2prod} .

\begin{eqnarray*}
& &\Lambda_{(11,6,6)/(5,5)}(x) \Lambda_{(6,6,1)/(5)}(x+7i) =
    \Lambda_{(6,1)}(x+6i) \Lambda_{(11,11,6,6)/(10,5,5)}(x+i)    \\
& & +T_0(x\pm i) T_0(x\pm 6i) T_0(x+8i) T_0(x+13 i),
\end{eqnarray*}
where $T_7$ is replaced by $T_0$.

%
By using eqs. (\ref{defTthree}) and (\ref{Lambdadual2}) 
and taking account of normalization factors,  one finds
\begin{eqnarray*}
& &T^{(3)}(x-\dl i) T^{(3)}(x+\dl i) =
 T_0(x\pm \frac{1}{2}i)  T_0(x\pm \frac{1}{6}i) \\
& &+ 
 \frac{\phi(x- \frac{10}{6}i)}{\phi(x\pm \frac{8}{6}i) \phi(x\pm \frac{10}{6}i)}
 T^{(2)}(x) \Lambda_{(11,11,6,6)/(10,5,5)}(x+\frac{5}{6}i).  \\
\end{eqnarray*}
Then the equivalent statement to eq.(\ref{t7})  is,
$$
\Lambda_{(11,11,6,6)/(10,5,5)}(x+\frac{5}{6}i) =
\phi(x \pm \frac{8}{6}i) \phi(x- \frac{10}{6}i) T^{(4)}(x) T^{(7)}(x). 
$$
This is nothing but  Lemma \ref{huget4t7}.
One thus completes the proof of proposition. $\fsquare(0.2cm, )$

In next section, we summarize similar results 
for the dilute $A_6$ model without proof.

%
%
\section{The $T-$ system for the dilute $A_6$ model }\label{summary-A6}

We firstly comment on the "duality property" in  the dilute $A_6$ model,
$$
T_m(x) =T_{11-m}(x), \qquad m=0, \cdots, 11
$$
and $T_{12}(x)=T_{13}(x)=0$.
This is again compatible with the symmetry of functional relations
but is still a conjecture supported by numerics.

The Yangian representation theory asserts that irreducible modules of 
$Y(E_6)$ are made by tensoring minimal objects, $W_1(x)$ and $W_5(x')$.
On the other hand, we assume 
$$
T_1(x)=T^{(6)}(x)
$$ 
as QTM for $W_6(x)$.
Thus the situation is  similar to the $E_7$ case;
one must figure out eigenvalues
of QTMs associated to $W_1(x)$ and $W_5(x')$ independently from
the knowledge of the dilute $A_6$ model.
We conjecture that eigenvalues of QTMs for these are same and
its explicit form reads,
\begin{eqnarray}
& &T^{(1)}(x)=T^{(5)}(x) \np
& &=
		\frac{1}{2} \Bigl (
  \omega^3 \phi(x+\frac{3}{4}i) \frac{Q(x+\frac{3}{4}i)}
                                     {Q(x-\frac{1}{4}i)}      +
  \omega^2 \phi(x-\frac{5}{4}i) \frac{Q(x-\frac{5}{4}i)Q(x-\frac{7}{4}i)}
                                     {Q(x-\frac{1}{4}i)Q(x+\frac{5}{4}i)} \np
 & &+ \omega \phi(x+\frac{1}{4}i) \frac{Q(x+\frac{1}{4}i)Q(x-\frac{7}{4}i)}
                                     {Q(x-\frac{3}{4}i)Q(x+\frac{5}{4}i)} +
  \phi(x-\frac{7}{4}i) \frac{Q(x-\frac{7}{4}i)^2}
                                     {Q(x-\frac{3}{4}i)Q(x+\frac{3}{4}i)}  \np
& &+ \frac{1}{\omega} \phi(x-\frac{1}{4}i) \frac{Q(x-\frac{1}{4}i)Q(x+\frac{7}{4}i)}
                                     {Q(x+\frac{3}{4}i)Q(x-\frac{5}{4}i)} +
 \frac{1}{\omega^2} \phi(x+\frac{5}{4}i) \frac{Q(x+\frac{5}{4}i)Q(x+\frac{7}{4}i)}
                                     {Q(x+\frac{1}{4}i)Q(x-\frac{5}{4}i)} \np
& &+  \frac{1}{ \omega^3} \phi(x-\frac{3}{4}i) \frac{Q(x-\frac{3}{4}i)}
                                                    {Q(x+\frac{1}{4}i)}									 
		    \Bigr ).
\end{eqnarray}
The following relations hold in parallel to eqs.(\ref{tstsone})-(\ref{tstsfive}),
which  are key ingredients in the proof of the $T-$ system.
\begin{eqnarray}
(T^{(1)}(x))^2&=&T_2(x+i\frac{7}{4}) \label{t1t1one}\\
T^{(1)}(x+i\frac{1}{4}) T^{(1)}(x-i\frac{1}{4}) &=&
 \frac{1}{2} \Bigl( T_4(x) +T_0(x) \Bigr) \label{t1t1two} \\
T^{(1)}(x+i\frac{1}{2}) T^{(1)}(x-i\frac{1}{2}) 
  &=&\frac{1}{2} T_5 (x+i\frac{7}{4})  \label{t1t1three} \\
T^{(1)}(x+i\frac{3}{4}) T^{(1)}(x-i\frac{3}{4}) &=&
T_1(x)+\phi(x) T^{(1)}(x+ i\frac{7}{4}).
\label{t1t1four} 
\end{eqnarray}
(We omit relations obtained by $x\rightarrow x+\frac{7}{4}i$.)

Let other QTMs be
\begin{eqnarray*}
T^{(2)}(x)&=& T^{(4)}(x)  =
T^{(1)}(x-\frac{1}{4}i) T^{(1)}(x+\frac{1}{4}i)-T_0(x)   \\
T^{(3)}(x)&=&\frac{1}{\phi(x+i\frac{7}{4}) }
\Bigl (
 T_1(x-\frac{1}{4}i) T_1(x+\frac{1}{4}i)-T_0(x\pm\frac{3i}{4}) \Bigr), 
\end{eqnarray*}
then the following $T-$ system is valid.

\begin{prop}
\begin{eqnarray}
T^{(1)}(x- \dle6i) T^{(1)}(x+\dle6 i) &=& T_0(x)+T^{(2)}(x), \label{e6t1}\\
T^{(5)}(x-\dle6 i) T^{(5)}(x+\dle6 i) &=& T_0(x)+T^{(4)}(x),  \label{e6t5}\\
T^{(6)}(x-\dle6 i) T^{(6)}(x+\dle6 i) &=&
       T_0(x\pm\frac{3i}{4})
       +\phi(x+i\frac{7}{4}) T^{(3)}(x),      \label{e6t6}\\
T^{(2)}(x-\dle6 i) T^{(2)}(x+\dle6 i) &=&
  T_0(x\pm\frac{1}{4}i)+
 T^{(1)}(x) T^{(3)}(x),    \label{e6t2} \\
T^{(3)}(x-\dle6 i) T^{(3)}(x+\dle6 i) &=&
T_0(x )  T_0(x\pm\frac{1}{2}i)
+ (T^{(2)}(x))^2 T^{(6)}(x) ,  \label{e6t3}\np
T^{(4)}(x-\dle6 i) T^{(4)}(x+\dle6 i) &=&
  T_0(x\pm\frac{1}{4}i)+
 T^{(3)}(x) T^{(5)}(x).    \label{e6t4} \np
\end{eqnarray}
\end{prop}
First three relations are trivial re-writings of definitions.
Last three equations need nontrivial proof which we omit for brevity.

%
\section{The $Y-$systems and $E$ type  Thermodynamic Bethe Ansatz}\label{Ysys}

We define $Y-$ functions by combinations of $T^{(a)}$s and
transform the $T-$ systems into equivalent but
desired forms.
In the case of the  dilute $A_4$ case, they explicitly read,
\begin{df}
\begin{eqnarray*}
Y^{(1)}(x) &:=& \frac{\phi(\renx+\frac{10}{6}i)T^{(2)}(\renx)}
                           {T_0(\renx\pm \frac{5}{6}i) }  \qquad
Y^{(2)}(x) :=\frac{T^{(1)}(\renx) T^{(3)}(\renx)}
                   {T_0(\renx) T_0(\renx\pm \frac{4}{6}i)}\\   
Y^{(3)}(x) &:=& \frac{T^{(2)}(\renx) T^{(4)}(\renx) T^{(7)}(\renx)}
                  {T_0(\renx \pm \dl i)T_0(\renx \pm \frac{1}{2}i)}   
                  \qquad
Y^{(4)}(x) :=\frac{T^{(3)}(\renx) T^{(5)}(\renx)}
               {T_0(\renx) T_0(\renx \pm \frac{1}{3}i)
                }\\
Y^{(5)}(x) &:=&\frac{T^{(4)}(\renx) T^{(6)}(\renx) }
                    {T_0(\renx\pm \dl i)}
         \qquad
Y^{(6)}(x) := \frac{T^{(5)}(\renx)}
              {T_0(\renx)} \\
Y^{(7)}(x) &:=& \frac{T^{(3)}(\renx) }{T_0(\renx\pm\frac{1}{3}i) }.
\end{eqnarray*}
\end{df}

Similarly  for the dilute $A_6$ model,
\begin{df}
\begin{eqnarray*}
Y^{(1)}(x) &=& Y^{(5)}(x):=\frac{T^{(2)}(\ren6x)}{T_0(\ren6x)}  \\
Y^{(2)}(x) &=& Y^{(4)}(x):=
   \frac{T^{(1)}(\ren6x) T^{(3)}(\ren6x)}
                   {T_0(\ren6x\pm \dle6 i)}\\   
Y^{(3)}(x) &:=& \frac{(T^{(2)}(\ren6x))^2 T^{(6)}(\ren6x)}
                  {T_0(\ren6x) T_0(\ren6x \pm \frac{1}{2}i)}    \\
Y^{(6)}(x) &:=& \frac{\phi(\ren6x +i\frac{7}{4}) T^{(3)}(\ren6x)}
              {T_0(\ren6x \pm\frac{3i}{4})}. \\
\end{eqnarray*}
\end{df}
Then  new sets of functional relations (the $Y-$ system) 
follow from the $T-$systems.
\begin{thm}
Functional relations among $Y-$ functions exhibit the $E_{6,7}$ structure
in the following form, 
$$
Y^{(a)}(x- i)Y^{(a)}(x+ i) = \prod_{b\sim a} (1+ Y^{(b)}(x)), 
\qquad a=1,\cdots, a_{max}.
$$
Here $a_{max}=6 (7)$ for the dilute $A_6$ ($A_4$ ) model, respectively.
We denote $a\sim b$ if $a$ and $b$ 
are adjacent nodes in the $E_6 (E_7)$  Dynkin diagram.
\end{thm}
This coincides with the $E_{6,7}$ case of the universal $Y-$ system
in \cite{AlZ}.

The derivation of TBA from the $Y-$ system
needs some information on the analytic structures of
$Y^{(a)}(x), 1+Y^{(a)}(x)$.
As stressed in the survey section,
only the $Y-$ system with nice analytic properties (ANZC)
 yields an explicit algorithm in the 
evaluation of free energy.
%

We employ numerical calculations for some fixed values of $N$ and $\beta$ for
this purpose.
This is relatively facile as  one has only to deal with 
the largest eigenvalue sector.  
Though we have performed  the numerical calculation for small values of $N$,
it already reveals intriguing patterns  for zeros of $T^{(a)}(x)$, which
are also observed for the dilute $A_3$ model.
Namely, imaginary parts of coordinates
of zeros show the remarkable coincidence with 
 exponents related to mass spectra in Table 1.
We state  it as a conjecture for {\sl arbitrary} $N$.
\begin{cj}
Zeros of $T^{(a)}$ distribute along  approximately on
 lines, $\Im x \sim \pm \dl (a_j+1)$ for the dilute $A_4$ model
and $ \pm \dle6 (a_j+1)  $ for the dilute $A_6$ model. 
The set $\{a_j \}$ agrees with $\{a \}$ for the particle $j$ in Table 1 (Table2).
\end{cj}
Therefore, we have a lemma parallel to the dilute $A_3$ case. 
\begin{lem}
Assume that the above conjecture is valid. 
Then $\widetilde{Y^{(a)}}(x)$ and  $1+Y^{(a)}(x)$ 
are Analytic, NonZero and have Constant asymptotic
behavior (ANZC) in  strips $\Im x \in [-1,1], [-0^+, 0^+]$,
respectively.
\end{lem}
$\widetilde{Y^{(a)}}(x)$ is defined by
$$
\widetilde{Y^{(a)}}(x)=
\left\{
\begin{array}{rl}
  Y^{(a)}(x) /\{\kappa(x+i(1+\tilde{u})) \kappa(x-i(1+\tilde{u}))\},&
           \quad \mbox{for $a=1 (6), u<0 $}  \\
  Y^{(a)}(x) \kappa(x+i(1-\tilde{u})) \kappa(x-i(1-\tilde{u})),&
           \quad \mbox{for $a=1 (6), u>0 $}  \\
  Y^{(a)}(x) ,& \quad \mbox{ otherwise }\\
\end{array} 
 \right.
$$
and $\tilde{u}=6u (4u)$ for the dilute $A_4$ ($A_6$) model.
The renormalization factor is given by 
$$
\kappa(x) =\Bigl( \frac{\vartheta_1(i\pi v/4, \tau')}
                   {\vartheta_2(i\pi v/4, \tau')} \Bigr )^{N/2}.
$$
where  $ \tau'=5\tau (\frac{7}{2}\tau)$ for the dilute $A_4$ ($A_6$) model.

The significance of the above property is clear when one  
 considers these relations (to be precise, logarithmic derivatives of them)
 in Fourier space, or "$k"$ space.
Cauchy's theorem assures that all quantities satisfy 
 algebraic equations at same $k$ , i.e., without mixing of modes.
Thus they can be solved in an elementary way.
We omit  the explicit procedure, for it has been given for other models
\cite{JKSfusion, KSS98, SuzE8}.
The resultant coupled integral equations read
\begin{eqnarray}
{\ln} Y^{(a)} (x) &= &- \epsilon \delta_{a,t} 
  \widetilde{\beta } s(x)
     +   {\cal C}_{a,b} * \ln (1+Y^{(b)}) (x),   \np
s(x) &=& \frac{\delta}{2\pi} \sum_n e^{ik_n x} \frac{1}{2 \cosh k_n},  \np
{\cal C}_{a,b}(x) &=&s(x)(2 I-{\cal C}^{\mathfrak{g}} )_{a,b},  
\end{eqnarray}
where $ t=1 (6),  \widetilde{\beta }=12 \pi \beta (8 \pi \beta) $, 
and  ${\cal C}^{\mathfrak{g}}$ denotes the Cartan matrix for $E_7$($E_6$).
$\delta$ also depends on whether we are dealing with the dilute $A_4$ model
or the dilute $A_6$ model through 
$\delta=\pi^2/(2 \tau')$ and $ k_n= n \delta$.
We also adopt the abbreviation, 
$A*B(x):= \int^{2 \tau'/\pi}_{-2 \tau'/\pi} A(x-x') B(x') dx'$.

This is nothing but the conjectured TBA for the $E_{6,7}$ RSOS model
at level 2 \cite{BazResProg}.

The free energy is expressed via $Y-$ functions with the aid of eq.(\ref{t1}).
We shall only give the result for $\epsilon=1$.
\begin{eqnarray}
-\beta f &=& -\beta e_0 -\widetilde{\beta} b_1*s(0) +s*\ln (1+Y^{(1)})(0) \np
e_0 &:=& 
 \left\{
    \begin{array}{rl}
	\lambda [ \ln(\vartheta_1(\pi/10) \vartheta_1(4\pi/10) )]'&
	          \quad \mbox {for $A_4$ }  \\
     \lambda [ \ln(\vartheta_1(\pi/7)/\vartheta_1(3\pi/7) )]'&
	          \quad \mbox {for $A_3$ }\\	 
     \end{array} 
 \right.    \np
\widehat{b_1}(x): &=& 
 \left\{
    \begin{array}{rl}
	\frac{\sinh 3x + \sinh 9 x}{\sinh 10x},  &
	          \quad \mbox {for $A_4$ }  \\
	\frac{\sinh 6 x}{\sinh 7x},  &
	          \quad \mbox {for $A_3$. }  \\
     \end{array} 
 \right. 
\end{eqnarray}

\section{Conclusion}\label{conclusion}

We have seen that the $E_{6,7,8}$ structure appears 
in the dilute $A_{6,4,3}$ model;
exponents of mass scale , zeros of QTMs and TBAs.
These results strongly support the underlying $E$ type symmetry in
the dilute $A_L$ model.

A Yangian analogue of Young tableaux arises in
proof of the $T-$ system. 
The  combinatorial aspects provide interesting problems on their own.
We thus believe that the subject is  worth of further research.

\section*{Acknowledgments}
The author thanks organizers and participants 
of the conference "Physical Combinatorics"
where he enjoyed lots of discussions.
He also thanks  V.V. Bazhanov, M. Jimbo, J. Shiraishi and O. Warnaar
for discussion and comments.

\section*{Appendix}
The RSOS weights for the dilute $A_L$ model is given by
\begin{eqnarray}
\raise 2mm \vtop{\hbox{$a$}\hbox{$a$}}
\,\framebox[0.4cm][c]{$u$} \,
\raise 2mm \vtop{\hbox{$a$}\hbox{$a$}}
&=& \frac{\theta_1(6-u)\theta_1(3+u)}{\theta_1(6)\theta_1(3)}- 
    \frac{\theta_1(u)\theta_1(3-u)}{\theta_1(6)\theta_1(3)} \times \np
& & \Bigl(
           \frac{S_{a+1}}{S_a} \frac{\theta_4(2a-5)}{\theta_4(2a+1)}
          +\frac{S_{a-1}}{S_a} \frac{\theta_4(2a+5)}{\theta_4(2a-1)}
    \Bigr ), \np
\raise 2mm \vtop{\hbox{$a\pm 1$}\hbox{$\quad a$}}
\,\framebox[0.4cm][c]{$u$} \>
\raise 2mm \vtop{\hbox{$a$}\hbox{$a$}}&=&
\raise 2mm \vtop{\hbox{$a$}\hbox{$a$}}
\> \framebox[0.4cm][c]{$u$} \>
\raise 2mm \vtop{\hbox{$a$}\hbox{$a\pm 1$}}
=\frac{\theta_1(3-u)\theta_4(\pm 2a+1-u)}{\theta_1(3)\theta_4(\pm 2a +1)},
\np
\raise 2mm \vtop{\hbox{$\quad a$}\hbox{$a\pm 1$}}
\,\framebox[0.4cm][c]{$u$} \>
\raise 2mm \vtop{\hbox{$a$}\hbox{$a$}}&=&
\raise 2mm \vtop{\hbox{$a$}\hbox{$a$}}
\> \framebox[0.4cm][c]{$u$} \>
\raise 2mm \vtop{\hbox{$a\pm 1$}\hbox{$a$}}
=\Bigl ( \frac{S_{a\pm 1}}{S_a} \Bigr )^{1/2}
\frac{\theta_1(u)\theta_4(\pm 2a-2+u)}{\theta_1(3)\theta_4(\pm 2a +1)},
\np
\raise 2mm \vtop{\hbox{$a$}\hbox{$a$}}
\> \framebox[0.4cm][c]{$u$} \>
\raise 2mm \vtop{\hbox{$a\pm 1$}\hbox{$a\pm 1$}}&=&
\raise 2mm \vtop{\hbox{$a\pm 1$}\hbox{$\quad a$}}
\> \framebox[0.4cm][c]{$u$} \>
\raise 2mm \vtop{\hbox{$a\pm 1$}\hbox{$\quad a$}}
=\Bigl ( \frac{\theta_4(\pm 2a +3)\theta_4(\pm 2a-1)}
              {\theta_4^2(\pm 2a+1)} \Bigr )^{1/2}
\frac{\theta_1(u)\theta_1(3-u)}{\theta_1(2)\theta_1(3)},
\np
\raise 2mm \vtop{\hbox{$a\pm 1$}\hbox{$\quad a$}}
\> \framebox[0.4cm][c]{$u$} \>
\raise 2mm \vtop{\hbox{$\quad a$}\hbox{$a\mp 1$}}&=&
\frac{\theta_1(2-u)\theta_1(3-u)}{\theta_1(2)\theta_1(3)},
\np
\raise 2mm \vtop{\hbox{$\quad a$}\hbox{$a\pm 1$}}
\> \framebox[0.4cm][c]{$u$} \>
\raise 2mm \vtop{\hbox{$a\mp 1$}\hbox{$\quad a$}}&=&
-\Bigl(  \frac{S_{a-1} S_{a+1}}{S_a^2} \Bigr )^{1/2}
\frac{\theta_1(u)\theta_1(1-u)}{\theta_1(2)\theta_1(3)},
\np
\raise 2mm \vtop{\hbox{$\quad a$}\hbox{$a\pm 1$}}
\> \framebox[0.4cm][c]{$u$} \>
\raise 2mm \vtop{\hbox{$a\pm 1$}\hbox{$\quad a$}}&=&
\frac{\theta_1(3-u)\theta_1(\pm 4a+2+u)}{\theta_1(3)\theta_1(\pm 4a+2)} \np
& &+
\frac{S_{a\pm 1}}{S_a}
\frac{\theta_1(u)\theta_1(\pm 4a-1+u)}{\theta_1(3)\theta_1(\pm 4a+2)},  \quad
\hbox{ for } \pm 4a+2 \ne 0, \np
 &=&\frac{\theta_1(3+u)\theta_1(\pm4 a-4+u)}{\theta_1(3)\theta_1(\pm4 a-4)} \np
 & &+
 \Bigl (
        \frac{S_{a\mp1}\theta_1(4)}{S_a \theta_1(2)}-
	    	  \frac{\theta_4(\pm 2 a-5)}{ \theta_4(\pm 2a+1)}
                      \Bigr )
		\frac{\theta_1(u) \theta_1(\pm 4a-1+u)}
		     {\theta_1(3) \theta_1(\pm 4a-4)},  \hbox{  otherwise }. \\
%
\end{eqnarray}
Here $\theta_{1,4}(x) =\vartheta_{1,4}(\lambda x, \tau)$, 
\begin{eqnarray*}
\vartheta_1(x,\tau) &=&2 q^{1/4}\sin x
\prod_{n=1}^{\infty}(1-2q^{2n} \cos 2x+q^{4n})(1-q^{2n}), \\
\vartheta_4(x,\tau) &=&
\prod_{n=1}^{\infty} (1-2q^{2n-1} \cos 2x+q^{4n-2})(1-q^{2n}),
\end{eqnarray*}
and $q=\exp(-\tau)$.
$\lambda$ is a parameter of the model specified in section \ref{review-model}  and 
$S_a$ denotes
$$
S_a=(-1)^a \frac{\theta_1(4a)}{\theta_4(2a)}.
$$

%
%

\end{document}